\documentclass[twocolumn]{aastex631} %

\DeclareRobustCommand{\VAN}[3]{#2}
\let\VANthebibliography\thebibliography
\def\thebibliography{\DeclareRobustCommand{\VAN}[3]{##3}\VANthebibliography}

\usepackage{graphicx}	
\usepackage{amsmath}	

\begin{document}

\title{Blowin' in the non-isothermal wind: core-powered mass loss with hydrodynamic radiative transfer} 

\shorttitle{Blowin' in the non-isothermal wind}
\shortauthors{Misener et al.}

\correspondingauthor{William Misener}
\email{wmisener@carnegiescience.edu}

\author[0000-0001-6315-7118]{William Misener}
\affiliation{Department of Earth, Planetary, and Space Sciences, The University of California, Los Angeles \\
595 Charles E. Young Drive East \\
Los Angeles, CA 90095, USA}
\affiliation{Earth and Planets Laboratory, Carnegie Institution for Science \\ 5241 Broad Branch Road NW \\ Washington, DC 20015, USA}

\author[0000-0001-6460-0759]{Matthäus Schulik}
\affiliation{Department of Earth, Planetary, and Space Sciences, The University of California, Los Angeles \\
595 Charles E. Young Drive East \\
Los Angeles, CA 90095, USA}
\affiliation{Imperial Astrophysics, Department of Physics, Imperial College London \\
Prince Consort Road \\
London SW7 2AZ, UK}

\author[0000-0002-0298-8089]{Hilke E. Schlichting}
\affiliation{Department of Earth, Planetary, and Space Sciences, The University of California, Los Angeles \\
595 Charles E. Young Drive East \\
Los Angeles, CA 90095, USA}

\author[0000-0002-4856-7837]{James E. Owen}
\affiliation{Imperial Astrophysics, Department of Physics, Imperial College London \\
Prince Consort Road \\
London SW7 2AZ, UK}
\affiliation{Department of Earth, Planetary, and Space Sciences, The University of California, Los Angeles \\
595 Charles E. Young Drive East \\
Los Angeles, CA 90095, USA}

\submitjournal{ApJ}
\received{May 24, 2024}
\revised{December 9, 2024}
\accepted{January 6, 2025}

\begin{abstract}
The mass loss rates of planets undergoing core-powered escape are usually modeled using an isothermal Parker-type wind at the equilibrium temperature, $T_\mathrm{eq}$. However, the upper atmospheres of sub-Neptunes may not be isothermal if there are significant differences between the opacity to incident visible and outgoing infrared radiation. We model bolometrically-driven escape using \textsc{aiolos}, a hydrodynamic radiative-transfer code that incorporates double-gray opacities, to investigate the process's dependence on the visible-to-infrared opacity ratio, $\gamma$. For a value of $\gamma \approx 1$, we find that the resulting mass loss rates are well-approximated by a Parker-type wind with an isothermal temperature $T = T_\mathrm{eq}/2^{1/4}$. However, we show that over a range of physically plausible values of $\gamma$, the mass loss rates can vary by orders of magnitude, ranging from $10^{-5} \times$ the isothermal rate for low $\gamma$ to $10^5 \times$ the isothermal rate for high $\gamma$. The differences in mass loss rates are largest for small planet radii, while for large planet radii, mass loss rates become nearly independent of $\gamma$ and approach the isothermal approximation. We incorporate these opacity-dependent mass loss rates into a self-consistent planetary mass and energy evolution model and show that lower/higher $\gamma$ values lead to more/less hydrogen being retained after core-powered mass loss. In some cases, the choice of opacities determines whether or not a planet can retain a significant primordial hydrogen atmosphere. The dependence of escape rate on the opacity ratio may allow atmospheric escape observations to directly constrain a planet's opacities and therefore its atmospheric composition.
\end{abstract} %




\section{Introduction}\label{sec:intro}
Small planets, with radii $R < 4 R_\oplus$, close to their stars, with orbital periods $P < 100$~d, are abundant \citep{Fressin13}. Transit surveys have revealed that around 50 percent of Sun-like stars have at least one such planet in orbit \citep[e.g.][]{Dattilo2023}. Furthermore, these planets are distributed bimodally in radius, with the smaller `super-Earths' separated by the `radius valley' at $\sim 1.8 R_\oplus$ from the larger `sub-Neptunes' \citep{Fulton17}. For the subset of these planets with measured masses, the radius valley is also reflected in bulk density. The smaller super-Earths have high densities consistent with a terrestrial composition, while `sub-Neptunes' have lower bulk densities requiring a volatile component \citep[e.g.][]{WeissMarcy14, Rogers2015}. The local minimum in the radius distribution also decreases as incident bolometric flux decreases \citep{vE18, FultonPetigura2018}, and weakly decreases as the the mass of the host star decreases \citep{FultonPetigura2018, VanEylen2021}.

These observations can be well-explained by the effects of hydrodynamic escape \citep{OwenWu13, Ginzburg18}. In this paradigm, rocky cores form while the natal disk was still present and accrete a few to 10 percent of their total mass in hydrogen gas. As the disk disperses, the removal of external pressure support and resulting adiabatic expansion of the gas removes some of the accreted envelope in a process termed `spontaneous mass loss' \citep{Ginzburg16} or `boil-off' \citep{OwenWu16}. Following the spontaneous mass loss phase, trans-sonic, hydrodynamic winds continue to remove the primordial hydrogen. Some planets retained primordial envelopes against atmospheric escape, becoming the observed `sub-Neptune' population with low bulk densities and large radii. Meanwhile, the less massive and closer-in planets were completely stripped, leading to the smaller `super-Earths' with bulk densities consistent with an Earth-like composition. Alternative explanations for the radius bimodality, including a dichotomy between rocky and water-rich interiors \citep[e.g.][]{Zeng2019, Madhusudhan20, BurnMordasini2024} or a bimodality created due to the late accretion of hydrogen from a gas-depleted nebula onto sub-Neptunes \citep{LeeConnors2021, LeeKaralis2022}, are the subject of ongoing work in the community. However, none have been as successful in explaining the observed demographics, such as the change in the location of the radius valley as functions of stellar mass and orbital period and the scarcity of hot Neptunes, as hydrodynamic escape \citep{Gupta20, RogersGupta2021}.

Hydrodynamic escape theories are separated into two categories distinguished by the heating mechanism of the upper atmosphere: photo-evaporation \citep[e.g.][]{SekiyaNakazawa1980, Yelle2004, Murray-Clay2009, OwenWu13} and core-powered mass loss \citep{Ginzburg16, Ginzburg18}. In the photo-evaporation framework, high-energy XUV radiation from the host star heats the upper atmosphere to many thousand kelvins, driving rapid escape. However, the XUV output of a typical star declines rapidly with time \citep{JacksonDavis2012}, limiting the most vigorous stripping to the first few hundred million years of a planet's existence for solar-mass stars. Meanwhile, under core-powered mass loss, the bolometric radiation of the host star maintains a temperature close to the equilibrium temperature in the outer atmosphere, leading to a slower outflow than in photo-evaporation. Atmospheric loss is sustained by the high heat capacity of the silicate core relative to the atmosphere. As the planet cools into space, the core, thermally coupled to the base of the atmosphere, resupplies heat into the envelope, preventing radius contraction and promoting atmospheric stripping for longer time spans. The resupply of energy from the hot interior is implicitly included in photo-evaporative models, but the effect is sufficient to create the radius valley through a bolometrically-heated wind alone. Both photo-evaporation and core-powered, i.e. bolometrically-driven, mass loss reproduce the aforementioned demographic observations \citep{OwenWu17,Gupta19,RogersGupta2021}. The two mechanisms are also not mutually exclusive but rather should occur together, potentially enhancing one another \citep{OwenSchlichting2024}. Whether photo-evaporative or core-powered mass loss drives the escape is determined by whether XUV radiation is absorbed interior to the sonic radius of a bolometrically-driven Parker wind \citep{BeanRaymondOwen2021, OwenSchlichting2024}.

For the purposes of planet evolution models, both photo-evaporative and core-powered mass loss rates are often estimated analytically. For photo-evaporation, the energy-limited approximation, as first formulated in \citet{SekiyaNakazawa1980} and \citet{WatsonDonahue1981}, is often applied, with thermal conductivity neglected for hot exoplanets \citep[e.g.][]{lammer2003, baraffe2004, Erkaev2007, LopezFortney2013, ChenRogers2016}. Meanwhile, core-powered mass loss is usually modeled as a trans-sonic Parker wind at the equilibrium temperature of the planet \citep{Ginzburg18,Gupta19,MS21}, which has a simple analytic mass loss rate \citep{Parker1958}. This approximation was used because core-powered mass loss relies on bolometric heating of the outer atmosphere, which is expected to produce a nearly isothermal outer atmosphere at the equilibrium temperature \citep[e.g.][]{Guillot2010, PisoYoudin2014}, appropriate conditions for a Parker-type wind model. However, while photo-evaporation has been subject to detailed hydrodynamic simulations which benchmark the analytic approximations \citep[e.g.][]{Murray-Clay2009, OwenJackson12, Salz2016, KubyshkinaFossati2018, KubyshkinaFossati2021, KrennFossati2021, CaldiroliHaardt2022, SchulikBooth2023}, bolometrically-driven core-powered mass loss has not yet been thoroughly modeled with a coupled hydrodynamic radiative-transfer model, which is the focus of this manuscript.

Importantly, the outer atmospheres of sub-Neptunes need not be isothermal. For example, if the opacity to the infrared outgoing radiation, $\kappa_\mathrm{P, therm}$ is not the same as the opacity to the incident visible radiation, $\kappa_{\mathrm{P}, \odot}$, then the upper atmosphere will have a more complex, non-isothermal temperature structure, which can produce temperatures both hotter and colder than the equilibrium temperature in different regions. This was first theorized to have important implications for the flared, upper atmospheres of protoplanetary disks due to favorable opacity functions of primitive dust
\citep{calvet1991, chianggoldreich1997, dullemond2001}
and can also occur in the upper atmospheres of planets, even when presumed to be dust-free 
\citep{HubenyBurrows2003, Guillot2010}. This difference in opacities between spectral bands is often quantified as a ratio 
\begin{equation}\label{eq:gamma}
    \gamma \equiv \kappa_{\mathrm{P}, \odot}/\kappa_\mathrm{P, therm}.
\end{equation}
These non-isothermal temperature structures in the optically thin regions of atmospheres have been extensively studied in exoplanetary contexts \citep[e.g.][]{BurrowsHubeny2007, FortneyLodders2008, ItoIkoma2015, MollierevanBoekel2015, WyttenbachEhrenreich2015}, due to their effects on observed transmission spectra and upper atmospheric chemical equilibrium. 
But by changing the density profile of the upper atmosphere, such non-isothermal profiles also affect mass loss rates \citep{SchulikBooth2023, Schulik2024moon}. Variations in $\gamma$ are expected across the diverse temperatures, pressures, and compositions of the sub-Neptune population. In this work, we investigate whether consideration of non-gray effects can affect the expected mass loss rates, and therefore the evolution, of planets undergoing core-powered mass loss. In doing so, we present the first test of core-powered mass loss using a hydrodynamic radiative transfer model.

This paper is structured as follows. We begin with an overview of the key physics of upper atmospheric temperature structure and its expected effect on mass loss rates in Section \ref{sec:overview}. In Section~\ref{sec:methods}, we present the atmospheric structure and evolution theory that form the basis of this work and describe our simulation methods. In Section~\ref{sec:results}, we present and analyze our atmospheric mass loss rate and evolution simulation results. In Section~\ref{sec:discussion}, we discuss our key assumptions and future possible directions, followed by a conclusion in Section~\ref{sec:conclusions}.

\section{Overview and Motivations} \label{sec:overview}
We first provide an overview of the key physical processes that determine, to first order, the upper atmospheric structures we analyze in this manuscript. Previous work has shown that using the double-gray approximation, the radiative atmospheres of planets can be characterized by two temperature regimes \citep[e.g.][]{Guillot2010}. In the bottom of this region, where the optical depth to outgoing thermal radiation, $\tau_\mathrm{P, therm} > 1$, the atmosphere reaches thermal equilibrium. Therefore the expected temperature in the inner region is close to the planetary equilibrium temperature: $T_\mathrm{in} \approx T_\mathrm{eq}/2^{1/4}$. This inner region temperature (not to be confused with the `internal temperature' representing the net outgoing heat flux, $T_\mathrm{int}$) is achieved in the limit that $\gamma \gg 1$ or $\gamma \ll 1$, while $T_\mathrm{in}$ is expected to be closer to $T_\mathrm{eq}$ when $\gamma \sim 1$ \citep{ParmentierGuillot2014}.

Meanwhile, we term the expected temperature in the outer atmosphere, where the optical depth to incident stellar radiation $\tau_{\mathrm{P}, \odot} < 1$, $T_\mathrm{out}$. In a fully gray atmosphere, the entire radiative region should be isothermal at the blackbody temperature, i.e. $T_\mathrm{out} \sim T_\mathrm{in} \sim T_\mathrm{eq}$. However, non-gray, i.e. wavelength-dependent, opacities can change the temperature profile. If the opacity to incoming stellar radiation is lower than the opacity to outgoing thermal radiation ($\gamma < 1$), then the upper atmosphere radiates thermal energy away more efficiently than it absorbs incident stellar energy, and the temperature will be colder than the deeper regions ($T_\mathrm{out} < T_\mathrm{in}$). Conversely, if the opacity to incoming stellar radiation is higher than the opacity to outgoing thermal radiation ($\gamma > 1$), then the upper atmosphere will absorb incident stellar light without being able to cool efficiently. Thus the outer atmosphere will be hotter than the deeper regions, causing a temperature inversion ($T_\mathrm{out} > T_\mathrm{in}$). Therefore, this outer, or skin, temperature scales with the ratio of the thermal and stellar opacities and can be expressed as $T_\mathrm{out} = T_\mathrm{eq} (\gamma/4)^{1/4}$ \citep{Guillot2010, SchulikBooth2023}. In either case, due to steady state disequilibrium between the internal energy of the gas and the radiation field, an atmosphere with $\gamma \neq 1$ can have an outer temperature that deviates from the equilibrium temperature.

Related to these $\tau = 1$ surfaces is the transit radius, $R_\mathrm{trans}$. The transit radius is the radius at which the chord optical depth to incident stellar radiation is equal to 1, i.e.:
\begin{equation}\label{eq:transit_integral}
    1 = 2 \int_{R_\mathrm{trans}}^{\infty} \rho(r) \kappa_{\mathrm{P}, \odot} \mathrm{d}s
\end{equation}
where $s = (r^2 - R_\mathrm{trans}^2)^{1/2}$ is the chord length.

\begin{figure*}
    \centering
	\includegraphics[width=0.8\textwidth]{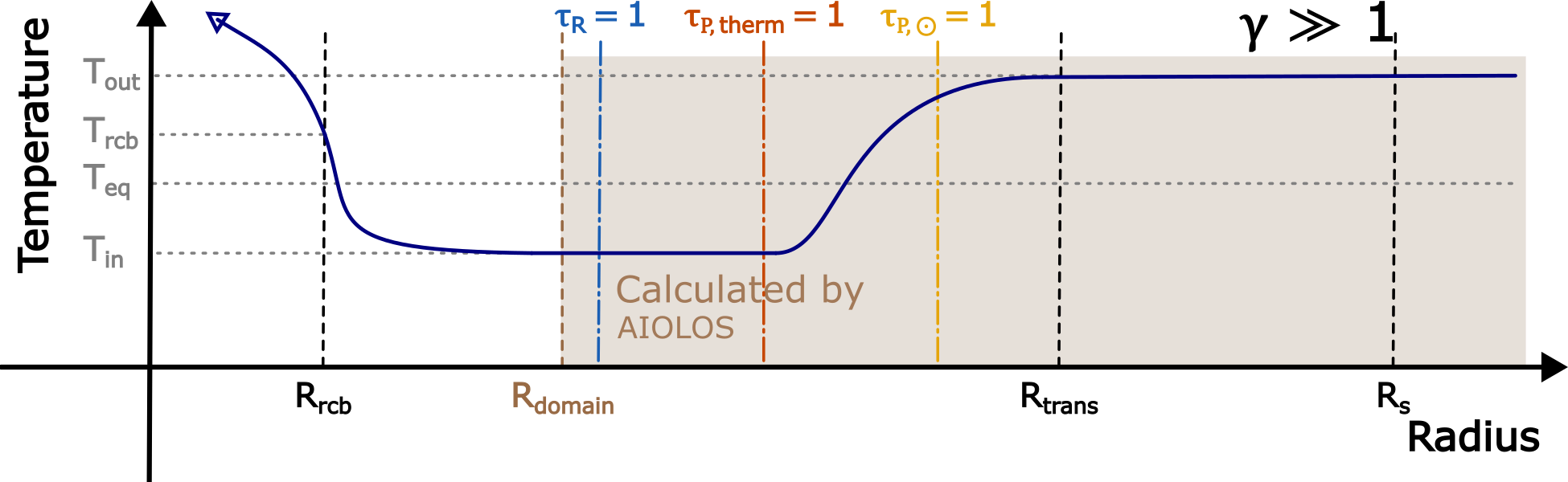}
	\includegraphics[width=0.8\textwidth]{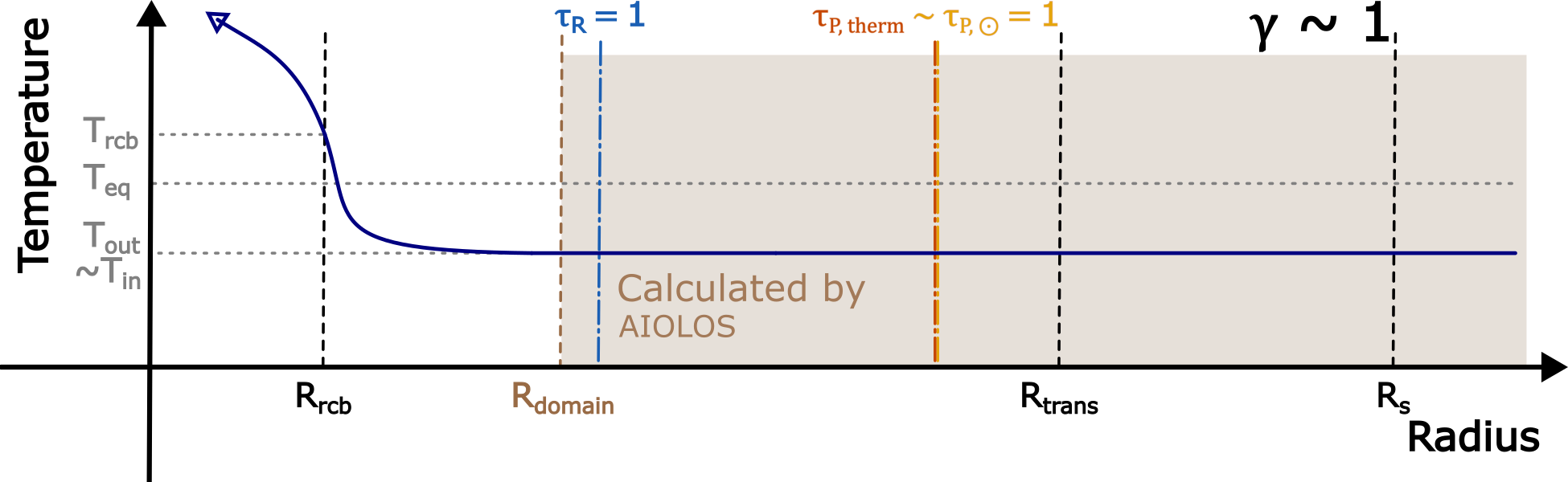}
	\includegraphics[width=0.8\textwidth]{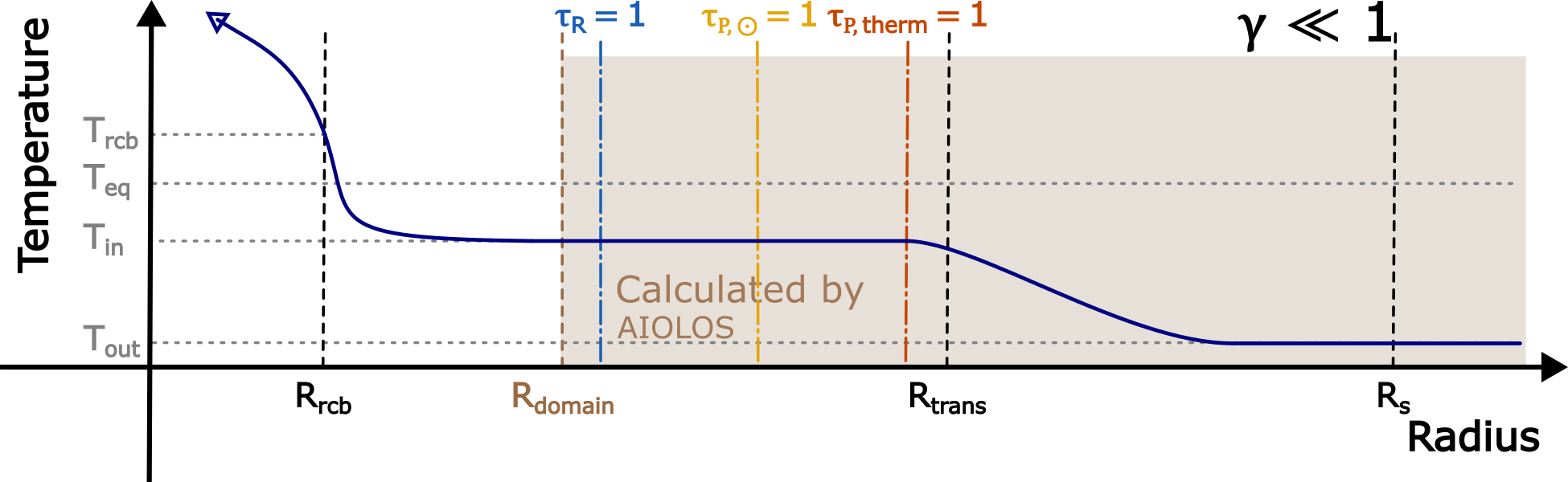}
    \caption{Schematic, not-to-scale representation of the typical temperature-radius profiles for different optical-to-infrared opacity ratios, $\gamma$. Three idealized cases $\gamma \gg 1$, $\gamma \sim 1$, and $\gamma \ll 1$ are illustrated on the top, middle, and bottom panels, respectively. Vertical black dashed lines show three key radii discussed in the text: the radiative-convective boundary radius $R_\mathrm{rcb}$, the transit radius $R_\mathrm{trans}$, and the sonic radius of the outflow $R_\mathrm{s}$. Additionally, three surfaces where the optical depth in different opacities is equal to 1 are shown by colored dash-dotted vertical lines: the thermal Rosseland optical depth, $\tau_\mathrm{R}$, in blue; the thermal Planck optical depth, $\tau_\mathrm{P, therm}$, in red; and the two-temperature stellar Planck optical depth, $\tau_{\mathrm{P}, \odot}$, in yellow. The tan shaded region denotes the domain of our \textsc{aiolos} hydrodynamic simulation domain, extending from its lower boundary $R_\mathrm{domain}$. Finally, horizontal gray dotted lines show four key temperatures discussed in the text: the equilibrium temperature $T_\mathrm{eq}$, temperature at the radiative-convective boundary $T_\mathrm{rcb}$, temperature in the inner radiative region $T_\mathrm{in}$, and temperature in the outer radiative region $T_\mathrm{out}$. The thermal profile strongly depends on the relative locations of the $\tau = 1$ surfaces, and therefore on the opacity ratio $\gamma$. Values of $\gamma > 1$ produce thermal inversions and high temperatures in the outer atmosphere, while values of $\gamma < 1$ produce cold outer atmospheres, significantly decreasing mass loss rates.}
    \label{fig:temp_diagram}
\end{figure*}

In Figure~\ref{fig:temp_diagram}, we summarize the key atmospheric physics described in this and the following sections. In each panel of the schematic, which is not to scale, we show example temperature profiles as functions of radii for three prototypical cases: an atmosphere with $\gamma \gg 1$, $\gamma \sim 1$, and $\gamma \ll 1$ from top to bottom. We show three key radii with dashed black vertical lines: the radiative-convective boundary (RCB) radius $R_\mathrm{rcb}$, the transit radius $R_\mathrm{trans}$, and the sonic radius of the outflow $R_\mathrm{s}$ (see Eq.~\ref{eq:sonic_radius} below). We also show with dash-dotted lines the radii at which the optical depth reaches unity for the three opacities that determine the structure of the atmosphere: the thermal Rosseland optical depth $\tau_\mathrm{R}$ in blue, the thermal Planck optical depth $\tau_\mathrm{P, therm}$ in red, and the two-temperature stellar Planck optical depth $\tau_{\mathrm{P}, \odot}$ in yellow. The domain of our hydrodynamic simulation is shown by the tan shading, extending outwards from its inner boundary $R_\mathrm{domain}$ (see Sec.~\ref{sec:extrapolation} below for details). Finally, we show the key temperatures defined in this section with gray horizontal dotted lines: the equilibrium temperature $T_\mathrm{eq}$, temperature at the radiative-convective boundary $T_\mathrm{rcb}$, temperature in the inner radiative region $T_\mathrm{in}$, and temperature in the outer radiative region $T_\mathrm{out}$.

In all three cases shown in Fig.~\ref{fig:temp_diagram}, at depth the temperatures are high and defined by an adiabat until the radiative-convective boundary is reached at some temperature $T_\mathrm{rcb}$, which, despite being shown as higher than is $T_\mathrm{eq}$, in reality may be higher or lower depending on the outgoing flux. In the radiative, optically thick regime, thermal diffusion determines the temperature gradient. This gradient begins equal to the adiabatic gradient but tends to zero as the radius increases and the pressure drops, per Eq.~\ref{eq:dlogTdlogP}, leading to a constant temperature close to $T_\mathrm{in}$.

From this point, the cases diverge. In the $\gamma \gg 1$ case, the $\tau_\mathrm{P, therm} = 1$ surface (red) is interior to the $\tau_{\mathrm{P}, \odot} = 1$ surface (yellow). In between these two, the atmosphere is optically thin to outgoing radiation but optically thick to incoming radiation. Therefore, most of the stellar energy is deposited in this region without it being able to efficiently cool. Therefore, the temperature increases with radius until $\tau_{\mathrm{P}, \odot} < 1$ and the atmosphere reaches a constant temperature at $T_\mathrm{out}$, which is higher than $T_\mathrm{eq}$. Conversely, in the $\gamma \ll 1$ case, the $\tau_{\mathrm{P}, \odot} = 1$ surface is interior to the $\tau_\mathrm{P, therm} = 1$ surface. In this case, the atmosphere is more efficient at cooling into space than heating, leading to cool outer temperatures and no heating beyond the equilibrium temperature. Thus, going outward, the atmospheric temperature is nearly isothermal near $T_\mathrm{in}$ until $\tau_\mathrm{P, therm} < 1$, at which point it begins declining to $T_\mathrm{out}$. Both of these temperatures, and thus the entire outer atmosphere, are lower than the equilibrium temperature. In the middle, $\gamma \sim 1$ case, $T_\mathrm{in} \sim T_\mathrm{out}$, so the optically thin region is nearly isothermal at a temperature slightly lower than the equilibrium temperature. In all cases, the transit radius $R_\mathrm{trans}$ is positioned close but slightly exterior to the $\tau_{\mathrm{P}, \odot} = 1$ surface, reflecting the increased path length of tangent photons compared to radial ones.

These changes in temperature produced by opacity alterations can affect the mass loss rates. One important change is to the density profile in the outer atmosphere. To first order, the density profile follows hydrostatic equilibrium, falling off exponentially with the lapse rate proportional to the temperature. Therefore, in regions where the temperature is lower than the equilibrium temperature, $T < T_\mathrm{eq}$, the density decreases more quickly with increasing radius than in an isothermal atmosphere at $T_\mathrm{eq}$. Conversely, in regions with $T > T_\mathrm{eq}$, the density falls off more slowly than in the isothermal case. Increasing the temperature also increases the sound speed of the gas, leading to a more rapid acceleration out of the planet's gravitational potential and a decrease in the sonic radius $R_\mathrm{s}$. To first order, these temperature effects lead to an increase in mass loss rate for planets with $\gamma \gg 1$ and a concomitant decrease in mass loss rate in $\gamma \ll 1$ cases. However, as we show in Section~\ref{sec:results}, the interaction between the various physical mechanisms described here can lead to additional complexity not captured by these simplified equations. We therefore employ a radiation-hydrodynamics code to model these atmospheres more self-consistently, which we describe in the following section.

\section{Methods} \label{sec:methods}
Our goal in this work is to simulate planet evolution under the core-powered mass loss framework of bolometrically-driven atmospheric escape. This requires connecting regions deep in the atmosphere, which are essentially hydrostatic, to hydrodynamic outflows. Since the evolutionary timescale of the deep interior is orders-of-magnitude longer than the outflow timescale (Myrs vs days), we choose to follow other evolutionary works by decoupling simulations of the evolution to outflow \citep[e.g.][]{baraffe2004, OwenWu13, LopezFortney2013, ChenRogers2016}. However, since bolometrically-driven outflows are more dependent on the interior properties than photoevaporation, we must be more careful in linking our outflow simulations to appropriate interior structures directly. We chose to do this using a combination of \textsc{aiolos}, a 1D radiation-hydrodynamics code \citep{SchulikBooth2023} \footnote{Publicly available at \url{https://github.com/Schulik/aiolos}}, and analytic structure models. In general, in the deep atmosphere energy is transported by convection. Eventually, densities and opacities become low enough that radiative diffusion dominates. In the outer atmosphere, the envelope becomes optically thin, and energy can escape directly into space. We expect the outer, optically thin atmosphere to have multiple competing physical processes, including absorption of radiation and hydrodynamic outflows, that we wish to investigate in detail in this work. Meanwhile, the deeper optically thick region of the envelope is very well approximated by the hydrostatic and radiatively diffuse limits. While \textsc{aiolos} can self-consistently calculate the atmospheric profile in all these regimes in both hydrostatic and hydrodynamic limits, its computational time increases with the optical depth. Therefore, to make the most efficient use of our computational resources, we choose to transition between a full computational solution from \textsc{aiolos} and an analytic profile once radiative diffusion becomes an appropriate approximation, i.e., at an optical depth $\sim 10$.

We begin this section by detailing the basic equations governing the model of atmospheric structure and evolution we employ in this work, including the ranges of atmospheric opacities we explore. We then detail our approach linking radiative-transfer hydrodynamic models of the upper atmosphere with evolution self-consistently.

\subsection{Atmospheric structure theory}\label{sec:method_structure_theory}
The basic picture of sub-Neptune composition and structure that we adopt in this work closely follows that of \citet{MS21}, which is one of a silicate-rich core surrounded by a hydrogen-rich envelope, motivated by extensive previous work \citep[e.g.][]{LopezFortney14, Ginzburg16, OwenWu17}. This composition is supported by comparison of mass loss models to the radius valley's location and its dependence on orbital period and stellar mass \citep{Gupta19, RogersOwen21}. In our model, an incompressible silicate core of mass $M_\mathrm{p}$ extends to a core radius $R_\mathrm{c}/R_\oplus \simeq (M_\mathrm{p}/M_\oplus)^{1/4}$ \citep{Valencia06}. Above the core, the hydrogen-rich envelope can be modeled to first order as a convective region at the base topped by a radiative region. The transition between these two regions occurs at the radiative-convective boundary radius, $R_\mathrm{rcb}$. In this work, we ignore the effects of compositional mixing between silicates and hydrogen that can lead to non-convective regions deep within sub-Neptunes \citep{MS22, Markham22} and to additional heat release due to condensation and demixing processes, including core formation and silicate rainout \citep[e.g.][]{VazanOrmel2024}.

In the convective region, the atmosphere follows an adiabatic profile, such that the density goes as:
\begin{equation}\label{eq:density_struc}
    \rho(r) = \rho_\mathrm{rcb} \bigg(1+\frac{R_\mathrm{B}'}{r} - \frac{R_\mathrm{B}'}{R_\mathrm{rcb}}\bigg)^{1/(\gamma_\mathrm{ad} - 1)}\mathrm{,}
\end{equation}
where $\rho_\mathrm{rcb} = \rho(r=R_\mathrm{rcb})$ is the density at the radiative-convective boundary and $\gamma_\mathrm{ad}$ is the adiabatic index of the atmosphere (not to be confused with the ratio of the visible and infrared opacities, which will be referred to as $\gamma$ throughout this work). We take $\gamma_\mathrm{ad} = 7/5$, appropriate for diatomic hydrogen. The so-called `modified Bondi radius' is defined for convenience as $R_\mathrm{B}' \equiv (\gamma_\mathrm{ad}-1)/\gamma_\mathrm{ad} \times G M_\mathrm{p} \mu/(k_\mathrm{B} T_\mathrm{rcb})$, where $G$ is the gravitational constant, $\mu = 2 m_\mathrm{p}$ is the molecular weight of H$_2$, $k_\mathrm{B}$ is the Boltzmann constant, and $T_\mathrm{rcb} = T(r=R_\mathrm{rcb})$ is the temperature at the radiative-convective boundary. Since the mass in the radiative region is typically negligible, the mass contained in the atmosphere, $M_\mathrm{atm}$, can be found by integrating Equation~\ref{eq:density_struc} over the convective region. 

Meanwhile, to calculate the planet's total energy, we follow \citet{MS21}. The total energy is the sum of the available energy in the core and atmosphere, $E = E_\mathrm{core} + E_\mathrm{atm}$. Since we take the core to be incompressible, the core's available energy is purely thermal: $E_\mathrm{core} = (1/(\gamma_\mathrm{core}-1)) M_\mathrm{p}/\mu_\mathrm{core} k_\mathrm{B} T_\mathrm{core}$, where $\gamma_\mathrm{core} \sim 4/3$ is the core's `adiabatic index', i.e. a representation of its specific heat capacity \citep{Scipioni17}, and $\mu_\mathrm{core} =60$~amu is the core's mean molecular weight. We assume the core is isothermal and thermally coupled to the base of the atmosphere, such that $T_\mathrm{core} = T(r = R_\mathrm{c})$. We neglect the energy released by differentiation in the interior, making our assumed core heat reservoir a conservative lower bound. However, the gravitational energy released by such processes will be at most the same order of magnitude as the formation energy of the planet, so our estimate is accurate to a factor of a few.

The atmosphere's energy is the sum of its gravitational potential and thermal energies:
\begin{equation}
    E_\mathrm{atm} = \int_{R_\mathrm{c}}^{R_\mathrm{rcb}} 4 \pi r^2 \left( -\frac{G M_\mathrm{c}}{r} + \frac{1}{\gamma-1} \frac{k_\mathrm{B} T(r)}{\mu} \right) \rho(r) \mathrm{d}r .
\end{equation}
As with the mass, the atmospheric energy is typically concentrated in the convective region \citep[e.g.][]{MS21}, so we do not consider the radiative region in its calculation. As the planet cools, the core's thermal energy decreases, and the atmosphere's total energy, typically negative, decreases as the atmosphere contracts. We define for convenience the energy ``available" for cooling, $E_\mathrm{avail}$, as the sum of the atmosphere's total energy, $-E_\mathrm{atm}$, and the core's thermal energy, $E_\mathrm{core}$:
\begin{equation}\label{eq:E_avail}
    E_\mathrm{avail} = E_\mathrm{core} - E_\mathrm{atm}.
\end{equation}
This quantity is useful for understanding the evolution of the planet in time (see Sec.~\ref{sec:methods_evol_theory} below). However, the numerical evolution uses the total energy, $E$.

For the atmospheres we consider, the radiative-convective boundary is highly optically thick. Therefore, the radiative diffusion approximation applies, and the temperature gradient just above $R_\mathrm{rcb}$ is:
\begin{equation}\label{eq:dlogTdlogP}
    \frac{\partial \ln T}{\partial \ln P} = -\frac{3 \kappa_\mathrm{R} P L}{64 \pi G M_\mathrm{p} \sigma T^4},
\end{equation}
where $R$, $P$, and $T$ are the local radius, pressure, and temperature, respectively, $L$ is the planet's luminosity, and $\kappa_\mathrm{R}$ is the Rosseland mean opacity. The radial pressure gradient in the radiative diffusion region is well-approximated by hydrostatic equilibrium, since the outflow velocities are negligible deep in the envelope:
\begin{equation}\label{eq:dPdr}
    \frac{\partial P}{\partial R} = -\frac{G M_\mathrm{p}}{R^2} \frac{\mu P}{k_\mathrm{B} T}.
\end{equation}

As the radius increases and the Rosseland mean opacity and pressures decrease, Eq.~\ref{eq:dlogTdlogP} tends to 0 and the temperature becomes roughly constant. In light of this, many atmospheric accretion and evolution models \citep[e.g.][]{LopezFortney14, OwenWu17}, including previous works examining core-powered mass loss \citep{Ginzburg16, Ginzburg18, Gupta19, MS21}, approximate the outer radiative region's temperature profile as isothermal at the planet's equilibrium temperature, $T_\mathrm{eq}$, such that $T_\mathrm{rcb} = T_\mathrm{eq}$. The equilibrium temperature depends solely on the stellar bolometric flux incident on the planet:
\begin{equation}\label{eq:Teq}
    T_\mathrm{eq} = \left(\frac{F}{4 \sigma}\right)^{1/4} ,
\end{equation}
where $F$ is the incident flux from the host star and $\sigma$ is the Stefan-Boltzmann constant. However, in this work we self-consistently connect our interior structure models to our simulations of the outer atmosphere, in which the temperature may not be equal to the equilibrium temperature. Therefore, we explicitly calculate the radiative temperature profile via Eq.~\ref{eq:dlogTdlogP} and find the RCB radius using the Schwarzschild criterion.

Equation~\ref{eq:dlogTdlogP} applies until radiative diffusion no longer dominates energy transport, i.e. the optical depth by the Rosseland mean opacity $\tau_\mathrm{R} < 1$. At this point, the temperature profiles depend strongly on the wavelength-dependent absorption and emission of light, and the physics described in Section~\ref{sec:overview} becomes paramount. To capture these complexities, in the results presented in this work we use a double-gray radiative-transfer hydrodynamic model to self-consistently calculate the temperature and density profiles above this point, which we describe in detail in Section~\ref{sec:methods_aiolos} below.

\subsection{Opacities}\label{sec:opacities}
To understand the basic physics of the problem, we use simplified opacities, using physically relevant values based on \citet{Freedman14}. Single-temperature Rosseland mean opacities, $\kappa_\mathrm{R}$, appropriate for thermal radiation in the optically thick diffusion limit, scale as:
\begin{equation}
    \kappa_\mathrm{R}=0.1 \left(\frac{\rho}{10^{-3}\mathrm{\,g\,cm^{-3}}}\right)^{0.6}\mathrm{\,cm^2\,g^{-1}}.
	\label{eq:Rosseland_scale}
\end{equation}

We use a constant value for the Planck single-temperature opacities, $\kappa_\mathrm{P, therm} = 7.5$\,cm$^2$\,g$^{-1}$, appropriate for thermal radiation in the optically thin regime. This opacity can, in reality, vary by a factor of 10 over a range of relevant temperatures and pressures \citep{Freedman14, MalyginKuiper2014}; we examine the possible impact of a different thermal Planck opacity in Section~\ref{sec:discussion_opacities} below. 
For the two-temperature opacity to incident stellar radiation, $\kappa_{\odot}$, we choose to use the Planck opacities, remaining consistent with \citet{Guillot2010}. It has also been suggested that the two-temperature Rosseland opacity, which leads to energy absorption primarily in the troughs of the opacity function is more appropriate to this problem \citep{Parmentier2015}. This choice of the best mean opacity could have a large impact on our results, since two-temperature Planck means are usually two to four orders of magnitude larger than the corresponding two-temperature Rosseland means \citep{Freedman14}. The ratio of opacities of interest, often termed $\gamma$, is then $\kappa_{\mathrm{P}, \odot}/\kappa_\mathrm{P, therm}$.

The opacities are a combination of broadband H absorption and line absorption by molecules in the infrared and alkali metals and oxides in the visible. Therefore, depending on a variety of atmospheric processes, including temperature and density, as well as composition, the value of the ratio could vary significantly from planet to planet. To understand the basic physics at play and make a first-order estimate of the magnitude of potential variation in mass loss rates, we adopt constant values of opacity for each run and iterate over a range of realistic opacities. Specifically, we test a range of values of the opacity to solar radiation $\kappa_{\mathrm{P}, \odot}$ ranging from 0.225 to 225\,cm$^2$\,g$^{-1}$, corresponding to opacity ratios between 0.03 and 30, which encompass the range of temperature, density, and compositions relevant to sub-Neptunes \citep{Freedman14}. In this way we remain agnostic to the actual opacity sources that cause the atmospheric effects we study here, and rather seek to quantify the magnitude of differences which are possible. In reality, using  wavelength-resolved opacities is optimal, because the exact line structure can have dramatic effects on the mass loss rate \citep{Schulik2024moon}. In the future, once these physical effects are better understood, real opacity models could be used to perform bespoke modeling of the evolution of individual planets. We discuss potential variations in opacity both radially within the atmosphere and through time in Section~\ref{sec:discussion_opacities}.

\subsection{Atmospheric evolution theory}\label{sec:methods_evol_theory}
In this work, we determine the atmospheric escape rate self-consistently from the \textsc{aiolos} hydrodynamic code. In the limit of an isothermal radiative region, as assumed in previous work examining core-powered mass loss, the mass loss rate, $\dot{M}_\mathrm{iso}$, can be solved analytically as a trans-sonic Parker wind \citep{Parker1958}:
\begin{equation}\label{eq:isothermal_loss}
    \dot{M}_\mathrm{iso} = 4 \pi R_\mathrm{s}^2 c_\mathrm{s} \rho_\mathrm{s} \approx 4 \pi R_\mathrm{s}^2 c_\mathrm{s} \rho_\mathrm{rcb} \exp{\left[2 - \frac{2R_\mathrm{s}} {R_\mathrm{rcb}} \right]} .
\end{equation}
In this equation, the isothermal sound speed $c_\mathrm{s} = (k_\mathrm{B} T/\mu)^{1/2}$, in which $T$ is the isothermal temperature and $R_\mathrm{s}$ is the resulting sonic radius:
\begin{equation}\label{eq:sonic_radius}
    R_\mathrm{s} = \frac{G M_\mathrm{p}}{2 c_\mathrm{s}^2} .
\end{equation}
In the right-most expression of Equation~\ref{eq:isothermal_loss}, this mass loss rate is written in terms of the radius, $R_\mathrm{rcb}$, and density, $\rho_\mathrm{rcb}$, at the base of the isothermal region, i.e. the radiative-convective boundary, above which the atmosphere is assumed to follow an exponential density profile \citep[e.g.][]{MS21}.

The planet also cools into space at a luminosity $L$. These rates allow the definition of two timescales that describe the evolution of small planets. One is the mass loss timescale, $t_\mathrm{loss} = M_\mathrm{atm}/\dot{M}$. The other is the cooling timescale $t_\mathrm{cool} = E_\mathrm{avail}/L$. When the mass loss timescale is shorter than the cooling timescale, the planet loses mass more quickly than it can cool and contract, leading to rapid atmospheric depletion. Conversely, if the cooling timescale is shorter than the mass loss timescale, the atmosphere will rapidly contract. Contraction brings the atmosphere deeper into the planet's gravitational potential and thus makes it more difficult to unbind, quickly throttling mass loss and leading to the preservation of the remaining atmosphere \citep{Gupta19, MS21}. 

\subsection{\textsc{aiolos} general setup}\label{sec:methods_aiolos}
We use \textsc{aiolos}, a 1D radiation-hydrodynamics code first presented in \citet{SchulikBooth2023}. We relegate most technical details to that paper, but we briefly restate the most important numerical approaches here. \textsc{aiolos} is a multispecies hydrodynamic code, solving one instance of the Euler system \citep{toro2009} per species via the HLLC Riemann solver \citep{toro1994}, using the \citet{kappeli2014} well-balancing scheme to stabilize hydrostatic regions against gravity. We only use one fluid in this work. The temperatures of our single gas species are computed in a double-gray approximation, with one radiation band representing the incoming and one the outgoing radiation. The incoming absorbed radiation is solved via the trivial solution of the Lambert-Beer law \citep{commercon2011, kuiper2013}, as we neglect scattering.
The outgoing, thermal radiation, whose mean photon field density is coupled to the gas temperature field, is treated numerically via flux-limited diffusion \citep{LevermorePomraning1981}
with the flux-limiter from \citep{Kley1989}, and the radiation matrix is solved similar to the methods presented in \citep{BitschCrida2013, lega2014}.

In this work, we consider an atmosphere of pure diatomic hydrogen gas with no incident UV flux. Dissociation and ionization of hydrogen are not important in the atmospheric regimes we consider here, so they are not included. However, consideration of these effects will be vital in future extensions of this work to incorporate photo-evaporation.

In all runs, we initialize the grid with an outer bound, which is fixed at $8 \times 10^{10}$\,cm\ $ \approx 125 R_\oplus$. This bound is well beyond the sonic radius of the outflows we consider. The inner bound determines the corresponding deep atmosphere, so we change it in each run as discussed in the next section. The grid is spaced at 500 cells per decade in the inner region, $r<8 \times 10^9$\,cm\ $ \approx 12.5 R_\oplus$, to resolve the shocks that sometimes develop early in the simulations, and 200 cells per decade in the outer region, where less resolution is required. The inner boundary is fixed at a constant density but allowed to evolve in temperature, while the outer boundary is open. We use a Courant number of 0.1 \citep{Courant1928, toro2009}.

\subsection{Run parameters}
For each run, we first specify the planet's mass $M_\mathrm{p}$ and semi-major axis $a$. We then specify the inner domain boundary, $R_\mathrm{domain}$. We also specify the internal flux at this inner boundary, which corresponds to an internal temperature $T_\mathrm{int}$ such that the planet's luminosity $L$ is given by
\begin{equation}\label{eq:internal_temp}
    L = 4 \pi R_\mathrm{domain}^2 \sigma T_\mathrm{int}^4 .
\end{equation}
To avoid numerical instabilities, we add this internal flux to \textsc{aiolos} by spreading it over the first three active cells in the domain. 

We fix the density at the bottom of the \textsc{aiolos} domain to $\rho_\mathrm{domain} \equiv \rho(R_\mathrm{domain}) = 10^{-5}$\,g\,cm$^{-3}$. We chose this value to correspond to a typical Rosseland optical depth between 1 and 10 at the bottom of the \textsc{aiolos} domain, as depicted on Fig.~\ref{fig:temp_diagram}. Such a value balances accuracy and computational speed: a denser domain greatly increases computational time and is well approximated by a hydrostatic diffusive profile, as described at the beginning of Sec.~\ref{sec:methods}.

We initialize the simulation as isothermal at the equilibrium temperature with a hydrostatic density profile. In all runs, we allow the code to evolve from its initial state, solving the equations of hydrodynamics and radiation transport, until it reaches a steady state outflow solution, i.e., the mass loss rate varies less than 5 per cent over the previous 25 percent of the simulation time. This typically takes on the order of $10^7$ to $10^8$\,s of simulation time for the regime we consider. In order to prevent instabilities, we ramp up the radiation gradually over the first $10^3$\,s. 

\subsection{Optically thick extrapolation}\label{sec:extrapolation}
To go from the bottom of the \textsc{aiolos} domain, which we choose to be weakly optically thick, to the deeper, adiabatic interior, we extrapolate following a diffusive temperature gradient. The bottom of the domain has radius $R_\mathrm{domain}$, density $\rho_\mathrm{domain} = 10^{-5}$\,g\,cm$^{-3}$, a temperature found by the the simulation, and an internal luminosity that corresponds to a specified internal temperature via Equation~\ref{eq:internal_temp} above. 

Beginning from the bottom of the \textsc{aiolos} domain, we increase the pressure by a small increment, $\Delta P = P/100$, and increment the temperature and radius according to Equations~\ref{eq:dlogTdlogP} and \ref{eq:dPdr} respectively. We recalculate the gradients, and continue this until the Schwarzschild criterion is met, i.e. $\partial \ln T/\partial \ln P = 2/7$. At this radius, the radiative-convective boundary, $R_\mathrm{rcb}$, we switch to an adiabatic gradient, calculated using the methods of \citet{MS21}. We integrate this adiabatic profile to infer an atmospheric mass $M_\mathrm{atm}$, often referred to as a fraction of the planet's total mass, $f \equiv M_\mathrm{atm}/M_\mathrm{p}$. In this way, we connect an \textsc{aiolos} boundary condition to the atmospheric evolution state it represents.

\subsection{Evolution}\label{sec:methods_evol}
To evolve the planets in time, we could in principle run full hydrodynamic radiative-transfer simulations with \textsc{aiolos} for every timestep to directly retrieve mass loss rates. However, such an approach would be very computationally expensive, and we expect the mass loss rates to vary smoothly over the phase space we consider. Therefore, to simulate planet evolution, we take the common approach \citep[e.g.][]{baraffe2004, OwenWu13, LopezFortney2013, ChenRogers2016} of making a grid of atmospheric structures calculated using hydrodynamic radiative-transfer simulations and interpolate over them.
We first run \textsc{aiolos} for a series of domain radii $R_\mathrm{domain}$, spaced 0.1 Earth radii apart, for each $\gamma$, $M_\mathrm{p}$, and $T_\mathrm{eq}$. Each run produces a particular atmospheric mass and energy, as well as a mass loss rate and luminosity. For sufficiently low luminosities (i.e., low internal temperatures), the temperature gradient at the base of the \textsc{aiolos} domain is nearly zero, so the profiles in the \textsc{aiolos} domain are, as expected, independent of the internal temperature. We verified empirically that so long as $\partial \ln T/\partial \ln P < 0.01$ at $R_\mathrm{domain}$, using a single run for all $T_\mathrm{int}$ yielded indistinguishable results from using a run with the actual $T_\mathrm{int}$, and so we employ one \textsc{aiolos} run for all possible $T_\mathrm{int}$ values in the work presented here to save computational time. We check \textit{a posteriori} that the gradient condition is always satisfied.

As discussed in Section~\ref{sec:intro}, as planets emerge from the disk they undergo rapid spontaneous mass loss due to the removal of pressure confinement. To simulate this effect, we choose an initial $R_\mathrm{rcb}$ by first thermally evolving the planets from an arbitrary size without any escape for $10^7$~yr, i.e. a typical disk lifetime. The resultant radius is virtually independent of the starting radius, and produces a planet with a cooling timescale of a few times $10^7$~yr, consistent with previous work \citep{OwenWu16, RogersOwen2024}. We choose a fiducial initial atmospheric mass, $M_\mathrm{atm} = 0.025 M_\mathrm{p}$, to represent a typical sub-Neptune \citep[e.g.][]{RogersOwen21}, but we vary this mass in Section~\ref{sec:params}. Starting from these initial $R_\mathrm{rcb}$ and $M_\mathrm{atm}$, we compute the time evolution by first finding the corresponding luminosity and \textsc{aiolos} mass loss rate. To do so, we interpolate over the grid of \textsc{aiolos} runs using the \textsc{scipy} \texttt{CubicSpline} function \citep{scipy} to find the mass loss rate as a function of arbitrary $R_\mathrm{domain}$.

We then subtract the mass and energy lost over a short time period $\Delta t= 0.01 \min{(t_\mathrm{cool}, t_\mathrm{loss})}$: $M_\mathrm{atm, new} = M_\mathrm{atm} - \dot{M} \Delta t$ and $E_\mathrm{new} = E - L \Delta t$. This evolution scheme is nearly identical to the methods used in \citet{MS21}, except with $\dot{M}$ taken from \textsc{aiolos} instead of an analytic Parker wind model. We then solve for the \textsc{aiolos} parameters $L$ and $R_\mathrm{domain}$ that correspond to the new mass and energy using the \textsc{scipy} \texttt{fsolve} root finding method \citep{scipy}. With the new atmospheric parameters in hand, we interpolate our grid of simulations to find the new mass loss rate and continue the iteration.

\section{Results}\label{sec:results}
In this section, we present the results of using \textsc{aiolos} derived outer atmosphere profiles for a range of visible-to-infrared opacity ratios. In Section~\ref{sec:gamma_results}, we compare instantaneous atmospheric profiles and demonstrate that incorporating more realistic opacities leads to non-isothermal profiles that can have a major effect on the atmospheric escape rate. Then, in Section~\ref{sec:evol_results}, we incorporate these profiles into planetary evolution calculations, finding that the fate of planetary atmospheres depends strongly on the visible-to-infrared opacity ratio $\gamma$ in the upper atmosphere.

\subsection{Sensitivity of mass loss rates to opacity}\label{sec:gamma_results}
\begin{figure*}
    \centering
    \includegraphics[width=0.7\textwidth]{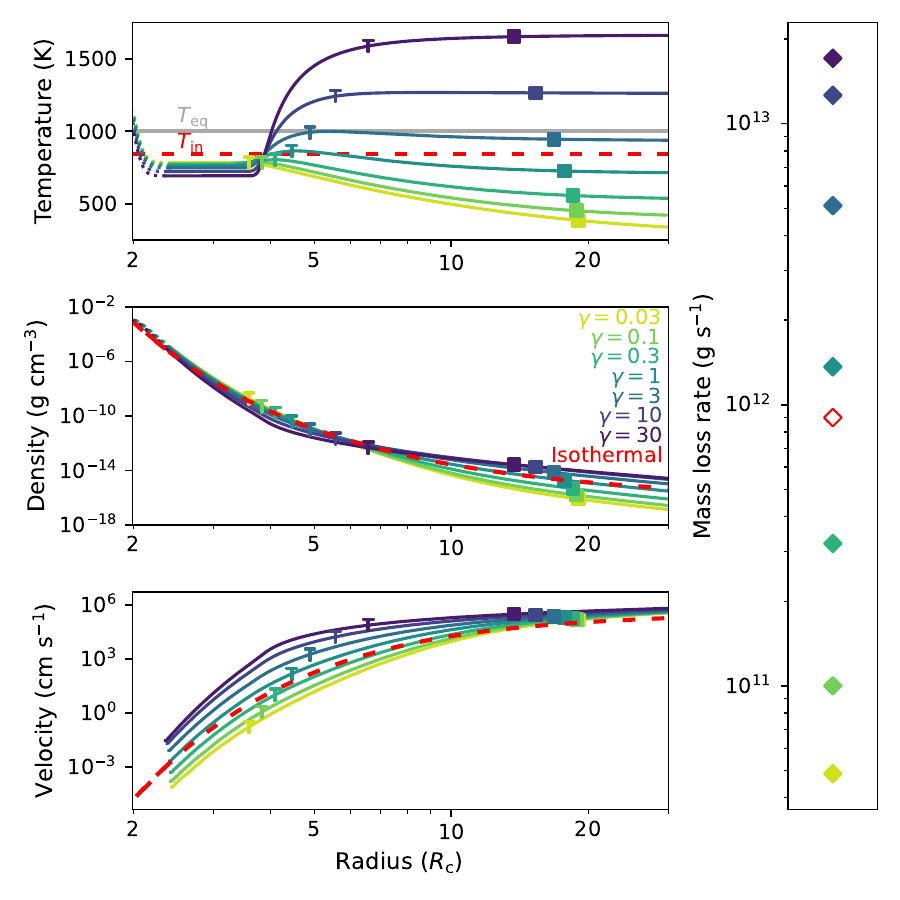}
    \caption{Temperature (top), density (middle), and velocity (bottom) profiles as functions of radius, in core radii $R_\mathrm{c}$, and corresponding mass loss rates (right) for seven values of the opacity ratio ranging from 0.03 to 30 (different colors from yellow to purple). All profiles have a fixed radiative-convective boundary location $R_\mathrm{rcb} = 2 R_\mathrm{c} = 1.91 \times 10^9$~cm, planet mass $M_\mathrm{p} = 5 M_\oplus$, atmospheric mass fraction $f=0.025$, and incident flux equivalent to an equilibrium temperature $T_\mathrm{eq} =  1000\,$K (shown by the gray horizontal line on the top panel). T-shaped markers represent the transit radius, while squares represent the sonic radius. Shown in red dashed lines are the equivalent profiles and mass loss rates for an isothermal radiative region at the analytic interior temperature, $T_\mathrm{in} = T_\mathrm{eq}/2^{1/4}$. Increasing the opacity ratio leads to different temperature and density profiles and therefore to mass loss rates that vary two orders of magnitude over plausible values. The mass loss rate for the $\gamma=1$ profile are similar to those for an isothermal outer atmosphere at $T_\mathrm{in}$.}
    \label{fig:profile_comp}
\end{figure*}

In Fig.~\ref{fig:profile_comp}, we compare the thermal (top left), density (middle left), and velocity (bottom left) profiles and mass loss rates (right) as functions of the ratio of the opacity to incident visible and thermal radiation, $\gamma$. Each color is a run with a value of this ratio ranging from 0.3 to 30, with $\gamma$ increasing as the colors get darker. All seven runs have a fixed radiative-convective boundary radius, $R_\mathrm{rcb}=2 R_\mathrm{c} = 1.91 \times 10^9$\,cm, a fixed initial atmospheric mass of $0.025 M_\mathrm{c}$, a planet mass $M_\mathrm{p} = 5 M_\oplus$, and a fixed incident flux equivalent to an equilibrium temperature of 1000\,K, which is shown by the gray horizontal line. For each $\gamma$ value, the solid line represents the output of the \textsc{aiolos} code, while the dotted line represents the semi-analytic extrapolation into the optically thick region described in Section~\ref{sec:extrapolation}. T-shaped markers represent the transit radius, $R_\mathrm{trans}$ (Eq.~\ref{eq:transit_integral}), while squares represent the sonic radius, $R_\mathrm{s}$ (Eq.~\ref{eq:sonic_radius}). The red dashed lines and red diamond represent results using an isothermal profile: a hydrostatic density profile with the same radiative-convective boundary radius and atmospheric mass, a constant temperature equal to $T_\mathrm{in} = T_\mathrm{eq}/2^{1/4}$, a Parker-type velocity profile following \citet{Cranmer2004}, and the corresponding mass loss rate following Eq.~\ref{eq:isothermal_loss}.

In general, the numerical thermal profiles reproduce the characteristics described in Section~\ref{sec:overview}, thus looking broadly similar to the schematic in Fig.~\ref{fig:temp_diagram}. As the opacity ratio $\gamma$ increases, the outer region, optically thin to incident radiation, heats up, with temperatures increasing by a factor of 7 over the studied 1000-fold opacity ratio increase. The outer temperatures range from well below the equilibrium temperature to well above it, and are well-predicted by the $T_\mathrm{out} = T_\mathrm{eq} (\gamma/4)^{1/4}$ relation of \citet{Guillot2010} and \citet{SchulikBooth2023}. Meanwhile the inner region, optically thick to outgoing radiation, reaches lower temperatures as $\gamma$ is increased, decreasing by 100\,K over the studied range. In all cases, the inner, optically thick radiative region is a few hundred kelvins cooler than the equilibrium temperature. This temperature is also lower by $\sim 70-200$~K than the analytic lower limit of $T_\mathrm{in}$ discussed in \citet{SchulikBooth2023}. The minimum temperature also increases monotonically with $\gamma$, rather than being highest at $\gamma \sim 1$. This is because the inner temperature approximation of \citet{ParmentierGuillot2014} was calculated assuming plane-parallel geometry. However, in small planets the scale height is a non-negligible fraction of the planet radius. Therefore, in our simulations the different surface areas of the absorption and emission surfaces lead to a decrease in temperature as $\kappa_{P, \odot}$ increases, since the same radiation is absorbed over a progressively larger area.

For $\gamma \sim 1$, the profile and resultant mass loss rate are well-approximated by an analytic, hydrostatic outer atmospheric profile at constant temperature $T_\mathrm{in}$ (in red). We therefore conclude that for a first-order analytic approximation of core-powered mass loss rates that match our hydrodynamic simulations, $T_\mathrm{in} \equiv T_\mathrm{eq}/2^{1/4}$ is the most appropriate temperature to use in Equations~\ref{eq:isothermal_loss} and \ref{eq:sonic_radius}. We note that this results in a lower mass loss rate than was typically used in past approximations of core-powered mass loss, that assumed $T_\mathrm{eq}$ in the isothermal region \citep[e.g.][]{Ginzburg16, Gupta19, MS21}. However, this change ultimately does not affect the formation of the radius valley, as the precise mass loss rates are degenerate with the assumed initial core and envelope masses (see Section~\ref{sec:discussion_demo} for further discussion).

The increase in the outer temperature with $\gamma$ leads to an increase in the hydrodynamic mass loss rate. For the example shown in Fig.~\ref{fig:profile_comp}, we find that the mass loss rate increases about two orders of magnitude as $\gamma$ increases from 0.03 to 30. Values of $\gamma > 1$ lead to larger mass loss rates than the isothermal analytic value, while values of $\gamma < 1$ lead to smaller mass loss rates. These changes in mass loss rate are the consequence of the mass loss rate's exponential sensitivity to the sonic radius, $R_\mathrm{s}$ (Eq.~\ref{eq:isothermal_loss}), which is inversely proportional to the local temperature (Eq.~\ref{eq:sonic_radius}). As shown in Fig.~\ref{fig:profile_comp}, the temperature in the outer region increases with $\gamma$, as expected from the scaling of outer temperatures described in Sec.~\ref{sec:method_structure_theory}, $T_\mathrm{out} \propto \gamma^{1/4}$. Therefore, as $\gamma$ increases, the sonic radius decreases (top), and the density profiles fall off more slowly at large radii than for lower $\gamma$ values (bottom). These two effects combine to increase the density at the sonic radius by orders of magnitude as $\gamma$ increases, proportionally increasing the mass loss rate per Eq.~\ref{eq:isothermal_loss} (right). We describe this effect further using an analytic toy model in Appendix~\ref{sec:app_toy}.

However, this increase in the mass loss rate with $\gamma$ is offset by the low temperatures, and therefore rapid fall-off in density with radius, in the inner optically thick regions as $\gamma$ increases: the high-$\gamma$ profiles have the lowest temperatures and densities at intermediate radii ($\sim 3-6 R_\mathrm{c}$ in Fig.~\ref{fig:profile_comp}). Therefore, the mass loss rate does not increase very quickly as the outer temperature increases.

\begin{figure}
	\includegraphics[width=\columnwidth]{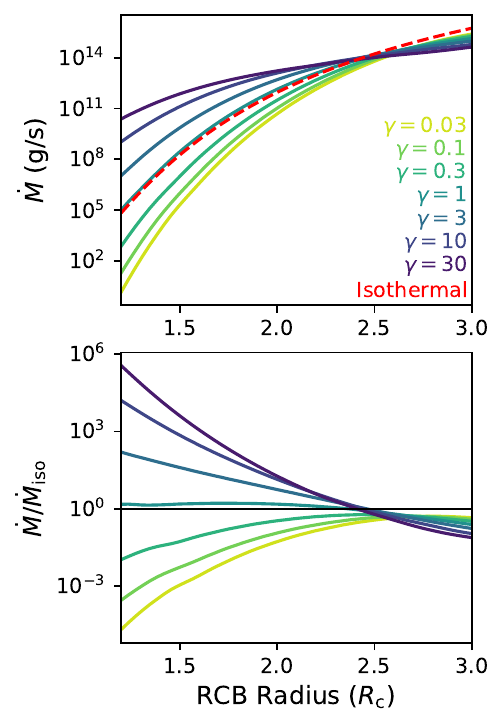}
    \caption{Mass loss rate derived from \textsc{aiolos} hydrodynamic simulations (top) and ratio of modeled mass loss rates to those found by an analytic isothermal Parker wind at $T_\mathrm{in}$(bottom), as functions of radiative-convective boundary radius, $R_\mathrm{rcb}$, in units of core radii $R_\mathrm{c}$. As in Fig.~\ref{fig:profile_comp}, the planets have a fixed atmospheric mass fraction, $f=0.025$, and incident flux equivalent to an equilibrium temperature $T_\mathrm{eq} =  1000\,$K. Each color from yellow to purple represents a different value of the visible-to-infrared opacity $\gamma$ ranging from 0.03 to 30, while the red line in the top panel represents the isothermal mass loss rate. Differences between $\gamma$ values are largest at low RCB radii, while the mass loss rates converge to a value close to the analytic value at high RCB radii.}
    \label{fig:Mdot_vs_RCB}
\end{figure}

The difference between different values of $\gamma$ depends on the radius of the planet. In the top panel of Figure~\ref{fig:Mdot_vs_RCB}, we show the mass loss rate as a function of radiative-convective boundary radius for different $\gamma$ values (in different colors from yellow to purple) taken from our hydrodynamic simulations. The red dashed line represents the mass loss rate given by an isothermal Parker wind at the analytic inner temperature $T_\mathrm{in}$. We show the ratio between the simulated mass loss rate and the analytic Parker wind rate in the bottom panel. At small RCB radii, the mass loss rate is a strongly increasing function of $\gamma$, as depicted in Fig.~\ref{fig:profile_comp}.
This is because at low RCB radii, the optically thin region is reached deep below the sonic point. Therefore, much of the atmosphere is nearly isothermal at $T_\mathrm{out}$, which depends on $\gamma$. For low values of $\gamma$, this means the density declines steeply before reaching the sonic radius, which is at cool temperatures and consequently low sound speeds. Conversely, for large values of $\gamma$, the density declines slowly and the sound speed at the sonic radius is high, leading to enhanced loss rates.

As the RCB radius increases, the dependence of the mass loss rate on $\gamma$ changes. For large RCB radii, the radius of the $\tau_\mathrm{P} = 1$ surfaces is sufficiently close to the sonic radius such that much of the radiative upper atmosphere is at $T_\mathrm{in}$ rather than $T_\mathrm{out}$. This change works to make the dependence on $\gamma$ weaker for large radii planets than for those with small radii: the mass loss rates converge to that given by an isothermal approximation at $T_\mathrm{in}$ for all values of $\gamma$. Additionally, for the lowest $\gamma$ runs, the mass loss rates increase more quickly with radius than in the analytic approximation. This is because the temperature in the outer regions declines slowly with radius, such that the temperature at the sonic radius can remain higher than $T_\mathrm{out}$. This enhances the mass loss rate both by increasing the sound speed and limiting the density drop-off. Meanwhile, in the high $\gamma$ cases, the interiors are even colder than the low $\gamma$ cases, leading to rapid density drops compared to the isothermal case as shown in Fig.~\ref{fig:profile_comp}. At small RCB radii, the outer high temperature regions allow the densities to `recover' and boost mass loss overall. But for large RCB radii there is not a large enough region of high temperature before the sonic point is reached, leading to mass loss rates that increase more slowly with radius than in the isothermal or low $\gamma$ cases. In the particular case shown in Figure~\ref{fig:Mdot_vs_RCB}, these effects are strong enough to reverse the dependence of the mass loss rate on $\gamma$, such that the lowest values of $\gamma$ have the highest loss rates, though all are within an order of magnitude of each other and the analytic approximations. This convergence of mass loss rates at large radii implies that the early evolution of planets, i.e. the spontaneous mass loss phase, is relatively insensitive to the opacity ratio $\gamma$ in the upper atmosphere.

\subsection{Changes to evolution}\label{sec:evol_results}
\begin{figure*}
	\includegraphics[width=\textwidth]{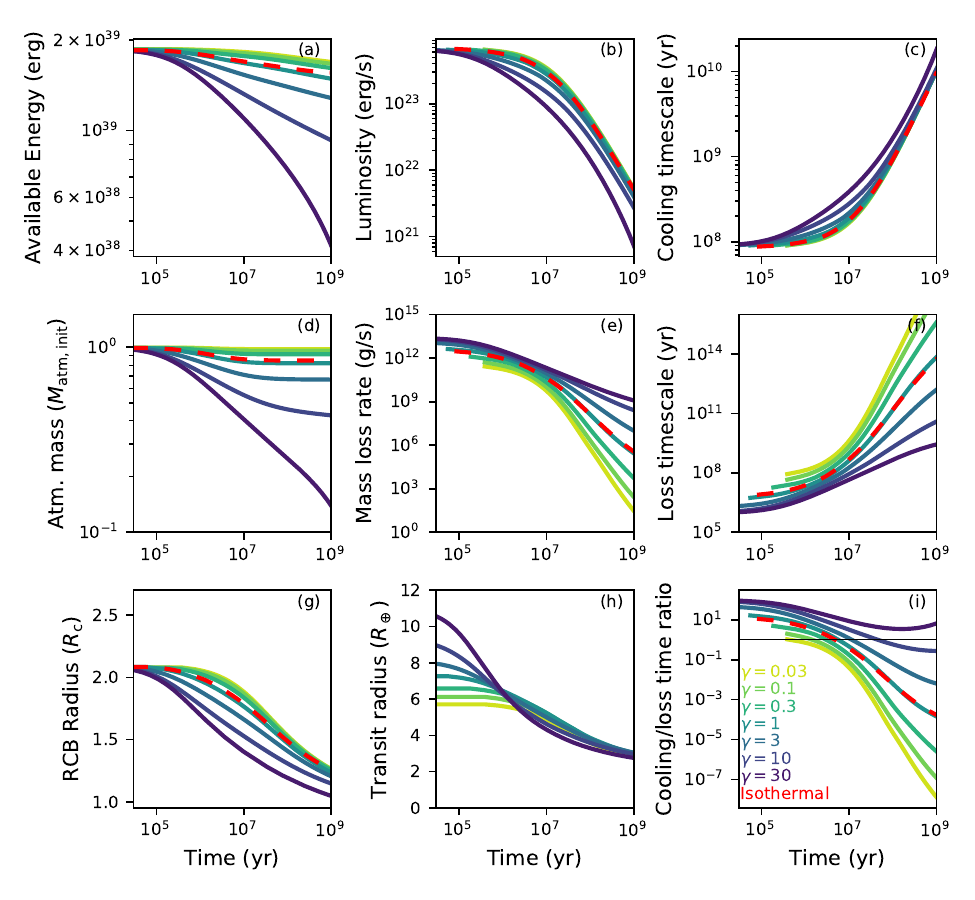}
    \caption{Evolution of planets with different initial equilibrium temperatures in time with mass loss rates derived from our \textsc{aiolos} hydrodynamics radiative-transfer simulations. In the top row we show the available energy, luminosity, and cooling timescale. In the middle row we plot atmospheric mass as a fraction of the planet's total mass, mass loss rate, and mass loss timescale. Finally, displayed on the bottom row are the RCB radius, transit radius, and ratio of the cooling timescale to the mass loss timescale. In all of these simulations, the planet mass $M_\mathrm{p} = 5 M_\oplus$, the initial RCB radius is 2.1 core radii, and the initial atmospheric mass fraction is 0.025 planet masses. Each color represents a different opacity ratio. A fully isothermal evolution, calculated using the methods described in \citet{MS21}, is shown with the red dashed lines. As the opacity ratio is increased, the planets lose more mass over time. The total mass lost can be either more or less than the isothermal prediction, depending on the value of the opacity ratio, with the $\gamma=1$ case very similar.}
    \label{fig:evol_compgamma}
\end{figure*}

The effects of different opacity ratios on mass loss rates combine to alter the mass and thermal evolution of small planets undergoing core-powered mass loss. In Fig.~\ref{fig:evol_compgamma}, we demonstrate how the evolution in time of a planet and the fate of its atmosphere can depend on the actual opacities in the outer atmosphere. We show the evolution in time of a planet with fixed planet mass $M_\mathrm{p} = 5 M_\oplus$, initial atmospheric mass fraction $f(t=0)=0.025$, and equilibrium temperature $T_\mathrm{eq} = 1000$~K. We vary these parameters and examine the effects in Section~\ref{sec:params}. Based on these initial conditions, we begin our evolution with an RCB radius $R_\mathrm{rcb}(t=0) = 2.1 R_\mathrm{c}$, which we determine by cooling the atmosphere without loss for $10^7$~yr as described in Sec.~\ref{sec:methods_evol}, to represent a typical sub-Neptune emerging from spontaneous mass loss. The different colors represent different choices of $\gamma$. We also show an evolution track of a planet with the same initial conditions but using an isothermal radiative region at $T_\mathrm{in}$ in red. This evolution was conducted using the methods described in \citet{MS21}, except that the `energy-limited' mass loss rate is ignored and only the Parker mass loss rate ($\dot{M}_\mathrm{B}$ in that work) is used. Additionally, motivated by the results presented in Section~\ref{sec:gamma_results}, this Parker mass loss rate has been modified to use $T_\mathrm{in}$ instead of $T_\mathrm{eq}$. On the top row of Fig.~\ref{fig:evol_compgamma}, as well as Figs.~\ref{fig:evol_compM}, \ref{fig:evol_compT}, and \ref{fig:evol_compf} below, we show the (a) available energy (as given by Eq.~\ref{eq:E_avail}), (b) planet luminosity, and (c) cooling timescale as functions of time. On the middle row, we show the (d) atmospheric mass as a fraction of the initial atmospheric mass, (e) mass loss rate, and (f) mass loss timescale. Finally, on the bottom row we show the (g) radiative-convective boundary radius in units of core radii, (h) transit radius in units of Earth radii, and (i) the ratio of the cooling timescale to the mass loss timescale.

Fig.~\ref{fig:evol_compgamma} shows that the evolution path of small planets that form with H/He envelopes are, in broad strokes, similar to those studied in detail in previous work such as \citet{MS21}. When small planets are young and have extended atmospheres, their mass loss timescales can be short, leading to rapid evolution in mass (panels (c) and (f)). Since the mass loss timescale is initially much shorter than the cooling timescale (panel (i)), these planets can lose significant fractions of their initial atmosphere to hydrodynamic escape (panel (d)). Cooling into space decreases the planet's total available energy (panel (a)), leading to atmospheric contraction (panels (g) and (h)). Significant escape is halted when the cooling timescale becomes shorter than the mass loss timescale (panel (i)). After this time, the planets cool and contract, usually leading to a rapid decrease in the mass loss rate (panel (e)) and increase in the mass loss timescale (panel (f)). However, for planets that reach very small radii, $R_\mathrm{rcb} \sim R_\mathrm{c}$, such the high-$\gamma$ runs represented by darker lines, rapid contraction is no longer possible, and the mass loss rate decreases more slowly. This can lead to an upturn in the ratio between the cooling and mass loss timescales (panel (i)), which otherwise decreases monotonically with time.

Despite the qualitative similarities, Fig.~\ref{fig:evol_compgamma} shows that different choices of opacities in the outer envelope can significantly alter the long-term evolution predicted by core-powered mass-loss. We find that as $\gamma$ increases, a planet loses more atmosphere over its lifetime (panel (d)). As shown in Section~\ref{sec:gamma_results}, this is due to the change in the temperature profile in the outer atmosphere, which changes the mass loss rate predicted for a given radius. Overall, different choices of upper atmospheric opacity can vary the final atmospheric mass by a factor of 10 or more. In addition, the apparent radius of these planets in transit also widely differs based on $\gamma$: the high $\gamma$ cases appear much larger in transit early in their evolution due to their large values of $\kappa_{\mathrm{P}, \odot}$ (panel (h)). 

\begin{figure}
	\includegraphics[width=\columnwidth]{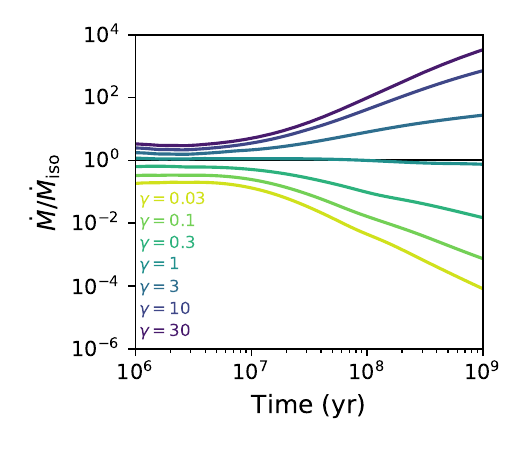}
    \caption{The ratio of the mass loss rates found using our hydrodynamic radiative-transfer model, \textsc{aiolos}, to those predicted analytically as functions of time since the beginning of the evolution. The evolution tracks are the same as those shown in Fig.~\ref{fig:evol_compgamma}. Each color represents a different value of $\gamma$. Initially, the simulated mass loss rates are very similar to those predicted analytically. Faster mass loss allows faster contraction for the higher $\gamma$ cases. While this lowers the mass loss rate, it brings the planet to radii for which the high temperatures of the outer radiative region greatly enhance loss (see Fig.~\ref{fig:Mdot_vs_RCB}). Therefore, as the planets contract, mass loss becomes more efficient for the high $\gamma$ cases relative to the isothermal case, leading to them sustaining higher mass loss rates over the course of the simulation. Conversely, the lower $\gamma$ cases become much less efficient at losing mass.}
    \label{fig:Mdot_vs_t}
\end{figure}

Additional perspective on these evolutionary differences can be gained from plotting the ratio between the mass loss rate found here and those predicted from the simulation that uses the analytic, isothermal mass loss rate as a function of time, which we show in Figure~\ref{fig:Mdot_vs_t}. Again, each color represents a different value of $\gamma$. Initially, the high $\gamma$ cases have moderately higher mass loss rates, but all the rates are fairly close to the isothermal value. However, these slightly higher mass loss rates allow the high $\gamma$  models to contract more quickly (Fig.~\ref{fig:evol_compgamma}, panel (g)). While their mass loss rate declines with time due to this contraction (see Fig.~\ref{fig:evol_compgamma}, panel (e)), the mass loss rate relative to the isothermal approximation increases. This is because as the planets contract, mass loss becomes relatively more efficient for the high $\gamma$ cases and much less efficient for low $\gamma$ cases, as is depicted in Fig.~\ref{fig:Mdot_vs_RCB}. This allows the $\gamma > 1$ runs to sustain mass loss rates above those of the isothermal simulation on billion year timescales, while the lower $\gamma$ runs essentially stop losing mass entirely, with the difference increasing with time. This result indicates that differences in $\gamma$ could control whether or not atmospheric escape from a small planet is detectable via transit spectroscopy (see Section~\ref{sec:discussion_obs} for further discussion).

\subsection{Variation with planet mass, equilibrium temperature, and initial atmospheric mass}\label{sec:params}
\begin{figure*}
	\includegraphics[width=\textwidth]{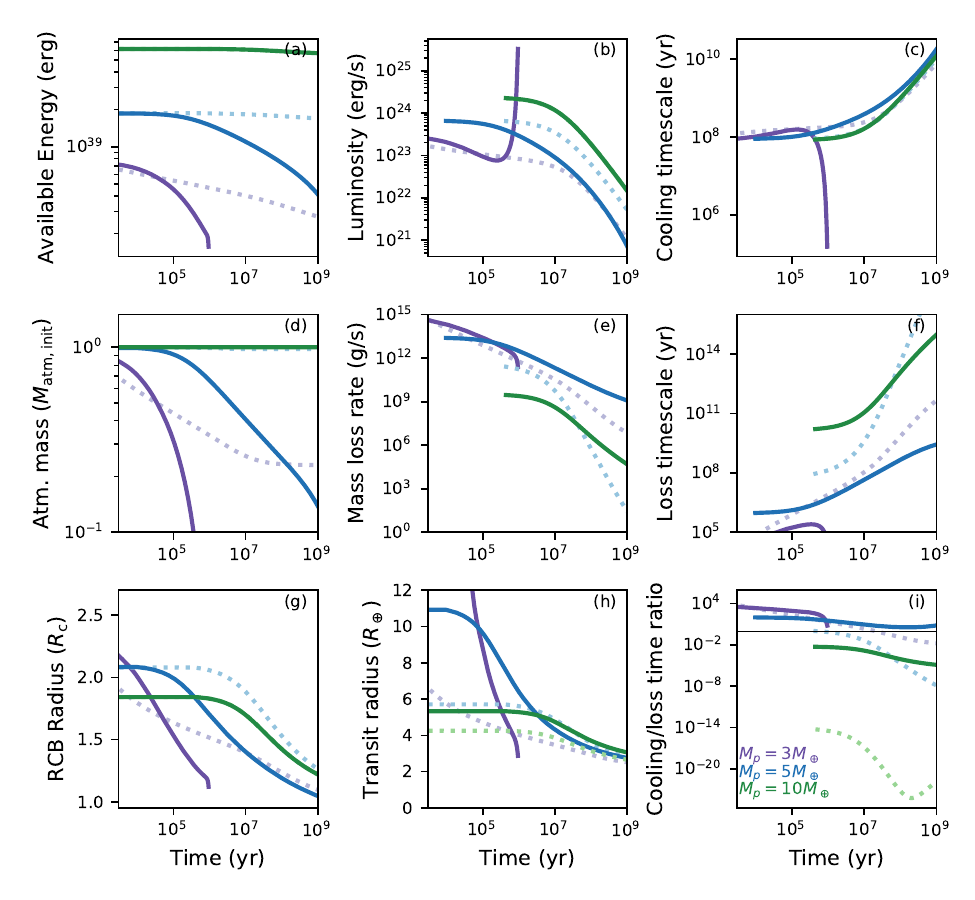}
    \caption{Evolution of planets with different initial planet masses in time with mass loss rates derived from our \textsc{aiolos} hydrodynamics radiative-transfer simulations. The panels are the same as in Fig.~\ref{fig:evol_compgamma}. Purple curves represent $3 M_\oplus$, blue curves represent $5 M_\oplus$, and green curves represent $10 M_\oplus$. The lightly shaded dotted line for each color represents a low opacity ratio, $\gamma=0.03$, and the darkly shaded solid line represents a high opacity ratio, $\gamma=30$, depicting the range of outcomes possible under different upper atmosphere opacities.}
    \label{fig:evol_compM}
\end{figure*}
The effects of changing the opacity ratio $\gamma$ vary depending on the planet mass, equilibrium temperature, and atmospheric mass. In Fig.~\ref{fig:evol_compM}, we show the variation in atmospheric evolution for different planet masses. All the models have the same equilibrium temperature, $T_\mathrm{eq} = 1000$~K and initial atmospheric mass fraction $f=0.025$, with the initial radii chosen as described in Section~\ref{sec:methods_evol}. The purple lines represent 3 $M_\oplus$, the blue lines represent 5 $M_\oplus$, and the green lines represent 10 $M_\oplus$. For each mass, the lighter colored, dotted line represents an evolution using a low opacity ratio, $\gamma=0.03$, while the darker colored, solid line represents an evolution using a high opacity ratio, $\gamma=30$. The panel meanings remain the same as in Fig.~\ref{fig:evol_compgamma}. For reference, the $5~M_\oplus$ planet tracks shown are the same as the $\gamma=0.03$ and $\gamma=30$ cases of Fig.~\ref{fig:evol_compgamma}, and these will also be shown in Figs.~\ref{fig:evol_compT} and \ref{fig:evol_compf} below.

As expected, the least massive planet loses the most mass (panel (d)). In fact, the high $\gamma$, $3~M_\oplus$ planet (dark purple) clearly undergoes rapid core-powered mass loss, in which contraction is slowed (panel (g)) by the heat released by the core, leading to a sharp increase in the luminosity and decrease in the cooling timescale with time (panels (b) and (c)), essentially becoming completely stripped. This is a marked difference from the low $\gamma$ $3 M_\oplus$ run, which loses only about 80\% of its initial atmospheric mass before cooling sufficiently (panel (d)). In this case, the choice of $\gamma$ is the difference between a planet retaining a 0.5\% weight percent, $\sim 10^4$~bar hydrogen-dominated envelope after core-powered mass loss and none at all. The 10 $M_\oplus$ planet loses essentially no mass at all irrespective of the choice of $\gamma$, as it begins its evolution in our simulations with its cooling timescale close to or shorter than its mass loss timescale (panel (i)).

\begin{figure*}
	\includegraphics[width=\textwidth]{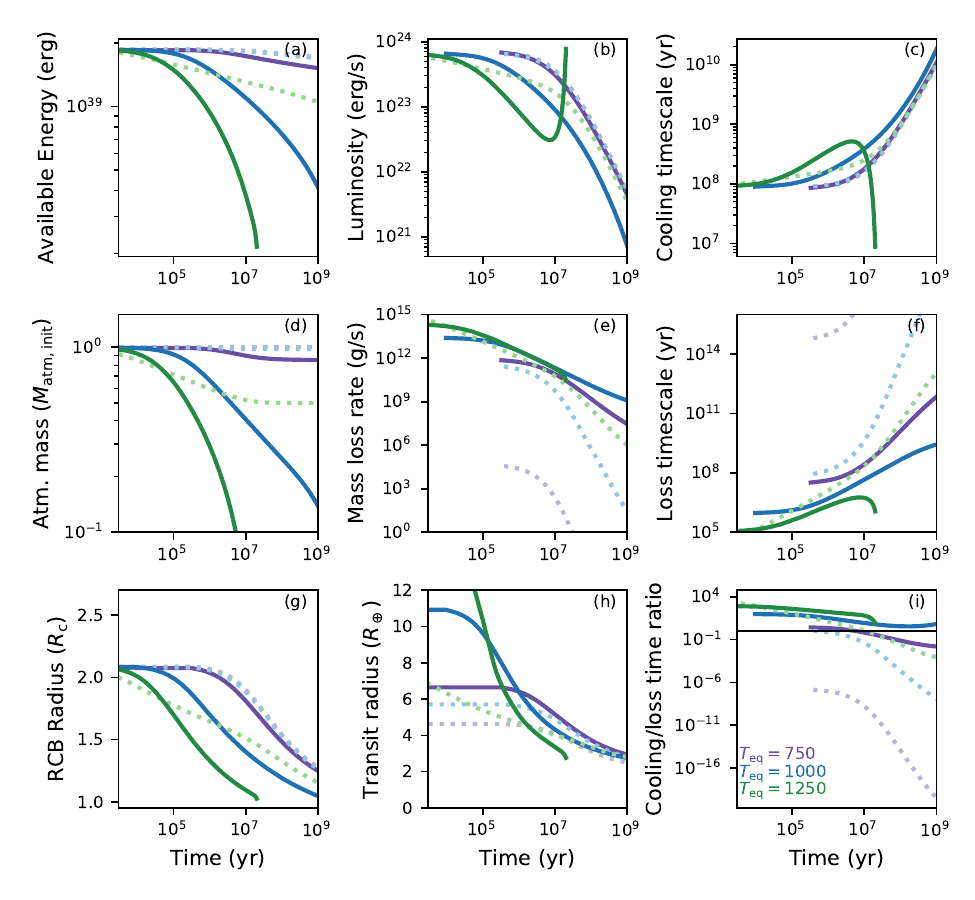}
    \caption{Evolution of planets with different initial equilibrium temperatures in time with mass loss rates derived from our \textsc{aiolos} hydrodynamics radiative-transfer simulations. The panels are the same as in Fig.~\ref{fig:evol_compgamma}. Purple curves represent $T_\mathrm{eq} = 750$~K, blue curves represent $T_\mathrm{eq} = 1000$~K, green curves represent $T_\mathrm{eq} = 1250$~K. The lightly shaded dotted line for each color represents a low opacity ratio, $\gamma=0.03$, and the darkly shaded solid line represents a high opacity ratio, $\gamma=30$, depicting the range of outcomes possible under different upper atmosphere opacities.}
    \label{fig:evol_compT}
\end{figure*}
In Fig.~\ref{fig:evol_compT}, we show the variation in atmospheric evolution for planets with equilibrium temperatures of 750, 1000 and 1250~K. All three models have the same planet mass, $5 M_\oplus$ and initial atmospheric mass fraction $f=0.025$. Their initial RCB radii are again chosen via  the method described in Section~\ref{sec:methods}. As expected, the hotter planets lose more mass, while the coolest planet loses very little: for $T_\mathrm{eq} = 750$~K, at least 80\% of the atmosphere is retained regardless of $\gamma$. The differences between choices of $\gamma$ are more notable for the hotter cases. For the hottest case shown in Fig.~\ref{fig:evol_compT}, $T_\mathrm{eq} = 1250$~K, the choice of $\gamma$ determines whether 50\% of the initial envelope or essentially none of it remains.

\begin{figure*}
	\includegraphics[width=\textwidth]{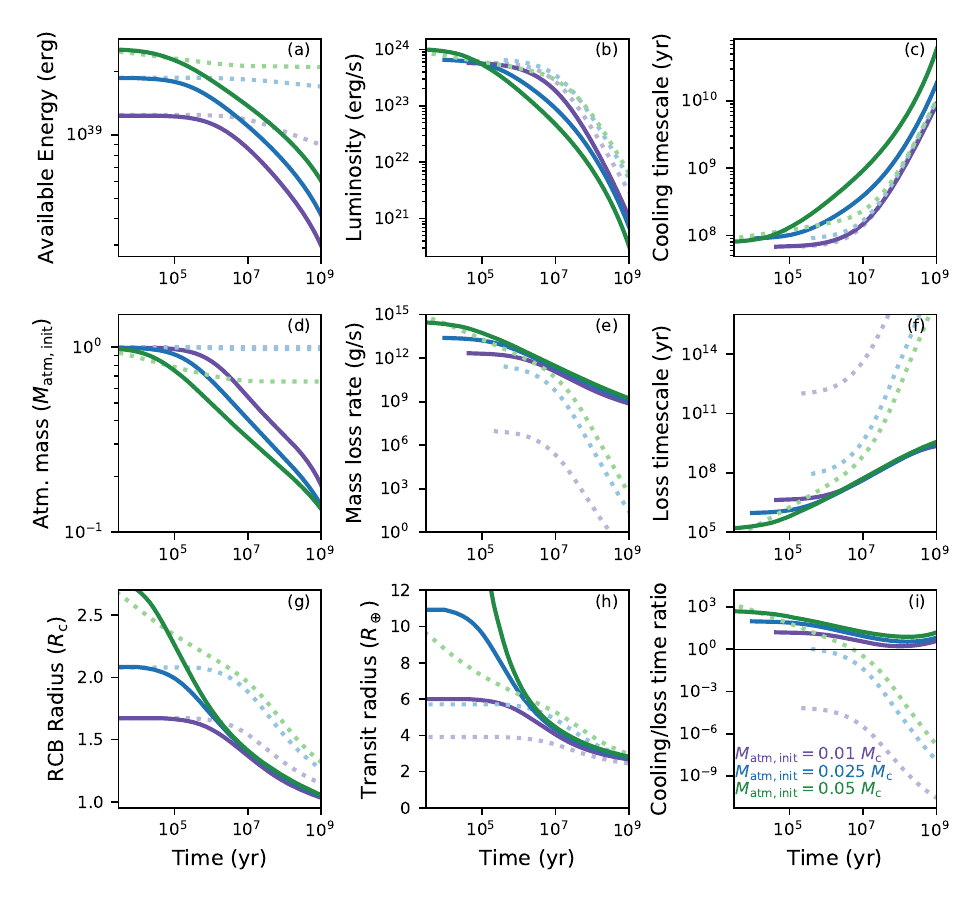}
    \caption{Evolution of planets with different initial atmospheric masses in time with mass loss rates derived from our \textsc{aiolos} hydrodynamics radiative-transfer simulations. The panels are the same as in Fig.~\ref{fig:evol_compgamma}. Purple curves represent $f = 0.01$, blue curves represent $f = 0.025$, and green curves represent $f = 0.05$. The lightly shaded dotted line for each color represents a low opacity ratio, $\gamma=0.03$, and the darkly shaded solid line represents a high opacity ratio, $\gamma=30$, depicting the range of outcomes possible under different upper atmosphere opacities.}
    \label{fig:evol_compf}
\end{figure*}
In Fig.~\ref{fig:evol_compf}, we show the variation in atmospheric evolution for planets with initial atmospheric mass fractions of 1, 2.5, and 5 percent. All three models have the same planet mass, $5 M_\oplus$ and equilibrium temperature $T_\mathrm{eq} = 1000$~K, with the initial RCB radius again chosen as described in Section~\ref{sec:methods_evol}. The more massive atmospheres begin larger and thus begin losing mass more quickly. In all three cases, planets lose at least five times as much mass over gigayear timescales in the $\gamma=30$ cases as in the $\gamma=0.03$ cases.

\section{Discussion}\label{sec:discussion}
\subsection{Implications for planet demographics}\label{sec:discussion_demo}
Using an isothermal approximation for the outer atmosphere, core-powered mass loss has been shown to be consistent with the observed radius valley \citep{Gupta19}. In this work, we demonstrate that changing the ratio between the thermal and stellar opacity, $\gamma$, in the outer atmosphere can change whether an atmosphere is retained or lost under the influence of core-powered mass loss. While it may be possible to use these large variations in mass loss rate to constrain the $\gamma$ distribution consistent with observed demographics, such an inference would be highly degenerate with other unknown population-level characteristics. Specifically, slight changes to the initial core mass distribution and the envelope masses and radii of the planets when they emerge from spontaneous mass loss could trade off with the altered mass evolution such changes to $\gamma$ would impart \citep[see, e.g.,][]{Gupta19, RogersOwen21}. It is also expected that $\gamma$ varies highly from planet to planet due to the diversity of compositions possible for sub-Neptune atmospheres. However, it may be possible to constrain past values of $\gamma$ for individual planets based on whether they retained a H/He envelope.

\subsection{Comparison to direct observations}\label{sec:discussion_obs}
A more direct test of the thermal structure of sub-Neptunes and its effects on planet evolution could come from atmospheric escape measurements. Signatures of escaping atmospheres have been observed directly for sub-Neptunes. Detections consistent with atmospheric escape for small exoplanets have been made using hydrogen Ly $\alpha$ 121~nm in the UV \citep[e.g.][]{KulowFrance2014, EhrenreichBourrier2015} and He 1083~nm in the NIR \citep[e.g.][]{MansfieldBean2018} lines (see \citet{DosSantos2023} for a recent review). If these planets are undergoing bolometrically-driven escape, measurements of the mass loss rate will constrain the temperature of the upper atmosphere, and therefore the atmospheric opacity ratio. However, the mass loss rate is difficult to constrain precisely from these measurements, due to degeneracies between escape velocity and ionization by the stellar wind \citep{OwenMurray-Clay2023, SchreyerOwen2024}. But using existing models, these measurements have been fit by isothermal mass loss rate estimates for sub-Neptunes as low as $10^9$~g s$^{-1}$ \citep{EhrenreichBourrier2015, MansfieldBean2018}. This measured rate is within the range of escape rates we predict on billion year timescales for some planet-$\gamma$ combinations, indicating that directly testing upper atmospheric structure with escape observations is within reach of existing telescopic capabilities. We facilitate comparisons to our results by tabulating the mass loss rates we obtain with \textsc{aiolos} as functions of planet mass, equilibrium temperature, value of $\gamma$, and transit radius and including them as machine-readable Supplementary Data. We describe these tables in Appendix~\ref{sec:app_tabulated}.

Spectroscopic observations with, for example, the \textit{James Webb Space Telescope} can probe the milli- to microbar temperature structure of small exoplanets. With sufficient precision, these can put direct constraints on the possible opacity values for those planets. In particular, inversions characteristic of large visible opacities lead to infrared molecular features in emission rather than absorption, as observed in some hot Jupiters \citep[e.g.][]{WyttenbachEhrenreich2015}.

\subsection{Opacity variations in space and time}\label{sec:discussion_opacities}
\begin{figure*}
    \centering
    \includegraphics[width=0.7\textwidth]{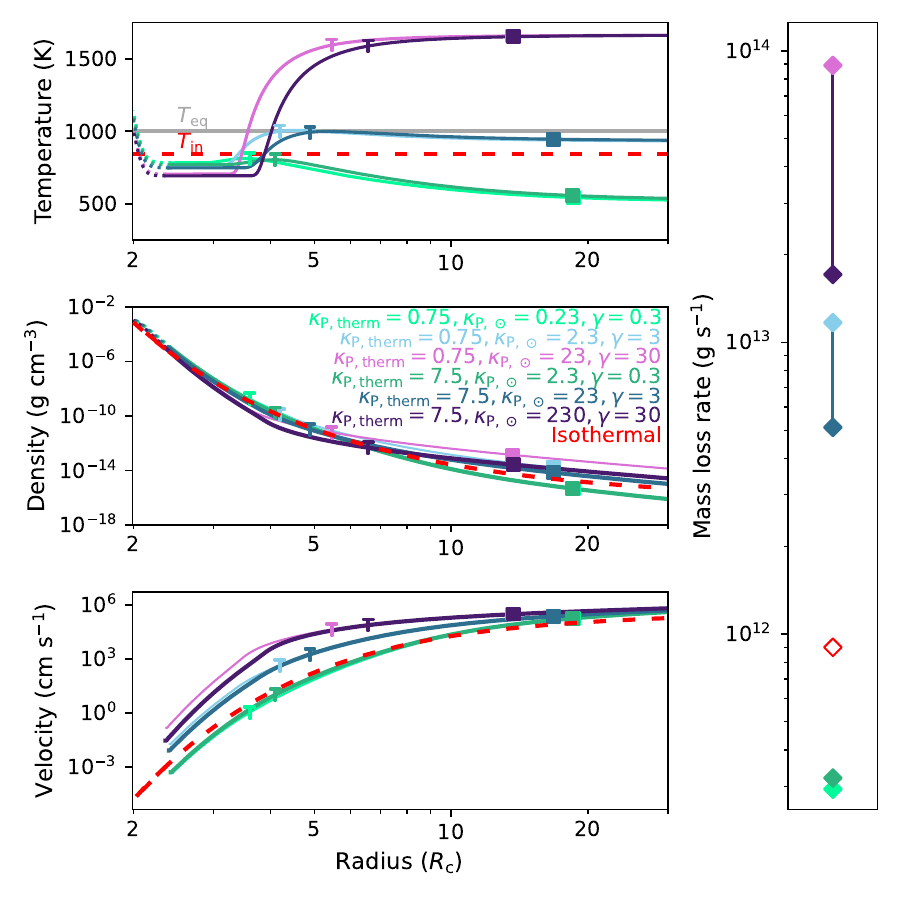}
    \caption{Temperature (top), density (middle), and velocity (bottom) profiles as functions of radius, in core radii $R_\mathrm{c}$, and corresponding mass loss rates (right) for six choices of Planck opacities. In the darker colors are three profiles with the default thermal Planck opacity presented in Section~\ref{sec:results}, $\kappa_\mathrm{P, therm} = 7.5$\,cm$^2$\,g$^{-1}$, and solar opacities such that the ratio between the two, $\gamma$ is 0.3 (green), 3 (blue), and 30 (purple). These profiles were previously shown in Fig.~\ref{fig:profile_comp}. The lighter shades of each color correspond to runs with the same $\gamma$ values, but all opacities lowered by a factor of 10, such that $\kappa_\mathrm{P, therm} = 0.75$\,cm$^2$\,g$^{-1}$. In the right panel, the mass loss rates for each particular value of $\gamma$ are linked by solid lines. All opacities listed in text on the figure are in cm$^2$\,g$^{-1}$.
    All profiles have a fixed radiative-convective boundary location $R_\mathrm{rcb} = 2 R_\mathrm{c} = 1.91 \times 10^9$~cm, atmospheric mass fraction $f=0.025$, and incident flux equivalent to an equilibrium temperature $T_\mathrm{eq} =  1000\,$K (shown by the gray horizontal line on the top panel). Shown in red dashed lines are the equivalent profiles and mass loss rates for an isothermal radiative region at the analytic interior temperature, $T_\mathrm{in} = T_\mathrm{eq}/2^{1/4}$. 
    As in Fig.~\ref{fig:profile_comp}, T-shaped markers represent the transit radius, while squares represent the sonic radius. Decreasing the opacities moves the $\tau = 1$ surfaces inward, and therefore the transitions between $T_\mathrm{in}$ and $T_\mathrm{out}$ move inward for each $\gamma$. For $\gamma > 1$, where $T_\mathrm{out} > T_\mathrm{in}$, this increases the mass loss rate, up to a factor of 5 for the $\gamma = 30$ case shown, while it lowers the mass loss rate for $\gamma < 1$. The value of $\gamma$ still sets the outer temperatures and therefore remains the main control on the mass loss rate.}
    \label{fig:profile_opacity_comp}
\end{figure*}

In this work, we chose a fixed Planck mean opacity to thermal radiation, $\kappa_\mathrm{P, therm} = 7.5$\,cm$^2$\,g$^{-1}$, varying the Planck mean opacity to incident stellar radiation, $\gamma_{\mathrm{P,\odot}}$ to achieve the range of $\gamma$ values studied. In reality, both mean opacities are functions of the density, temperature, and composition of the atmosphere. The stellar opacity varies more than the thermal opacity, with maximum and minimum tabulated values a factor of 1000 different over the relevant phase space of temperatures between 650 and 1500\,K, densities between $10^{-8}$ and $10^{-2}$\,g\,cm$^{-3}$, and solar metallicity per \citet{Freedman14}, compared to a factor of 10 difference in maximum and minimum values over the same range for thermal opacities. Higher opacity ratios than investigated here (i.e. $\gamma > 30$) are also predicted by some tabulations \citep{MalyginKuiper2014}. For these high $\gamma$ values, the mass loss rate will become energy-limited, saturating at a constant value \citep{Schulik2024moon}.

In order to ascertain that our results are independent to first-order of our exact choice of the thermal Planck opacity $\kappa_\mathrm{P, therm}$, we run simulations with a lower value, $\kappa_\mathrm{P, therm} = 0.75$\,cm$^2$\,g$^{-1}$, shown in Figure~\ref{fig:profile_opacity_comp}. Here we show the temperature, density, and velocity profiles as functions of radius, as well as the mass loss rate, similarly to Figure~\ref{fig:profile_comp}. The darker shades show three profiles with the default thermal Planck opacity presented in the results section, $\kappa_\mathrm{P, therm} = 7.5$\,cm$^2$\,g$^{-1}$, and solar opacities such that the ratio between the two, $\gamma$ is 0.3 (green), 3 (blue), and 30 (purple). The lighter shades show the same three choices of $\gamma$, but with both opacities decreased by a factor of 10. The mass loss rates for the same value of $\gamma$ are linked by solid lines in the right panel. We find that lower opacities move the $\tau = 1$ surfaces inward for the same $\gamma$. This moves the transition point from $T_\mathrm{in}$ to $T_\mathrm{out}$ inward. If $\gamma > 1$, then $T_\mathrm{out} > T_\mathrm{in}$, and the larger region of the atmosphere at higher temperatures leads to an increase in the mass loss rate. Conversely, if $\gamma < 1$, then the mass loss rate decreases. This increase can be up to a factor of 5 for $\gamma = 30$, but since $\gamma$ still determines the relative temperatures, the mass loss rates scale with $\gamma$ in a similar way. We therefore conclude that while the thermal structure, and therefore the mass loss rate, is sensitive to the choice of absolute opacities, such variation is second-order compared to the value of $\gamma$, which remains the main control on mass loss rate.

We have also assumed Planck opacities are constant as functions of planet radius and throughout a planet's evolution. However, these opacities depend on the composition of the atmosphere, which may vary significantly with both radius and time. From observations, sub-Neptunes appear to have a diversity of atmospheric compositions, with estimates varying from relatively low metallicities corresponding to solar composition or even more metal-poor \citep[e.g.][]{BaratDesert2024} to nearly even mixtures by mass of H/He and heavier species \citep[e.g.][]{BennekeRoy2024}. Theoretically, the chemical composition, and thus the opacity, of the upper atmosphere depends on a host of factors. Depending on the temperature profile, different gas species could be stable at chemical equilibrium, leading to variations in opacity and the opacity ratio with temperature \citep{Freedman14}. 
Vertical mixing could then make the composition of the upper atmosphere more uniform, causing apparent disequilibrium chemistry \citep{FortneyVisscher2020}. However, the vigor of such vertical mixing in sub-Neptune atmospheres, often parameterized as the diffusion parameter $K_{zz}$, is highly uncertain. Recent work has also shown the importance of chemical reactions with the interior for sub-Neptune atmospheric composition, since in these planets, the total mass is dominated by the silicate-rich interior \citep{SY22}. Considering these reactions leads to the potential for unexpected atmospheric constituents, such as silane \citep{MisenerSchlichting2023}, underscoring the inherent uncertainties in the opacity ratios typical of sub-Neptune upper atmospheres that await further investigation. The dependence of such reactions on the hydrogen content of the planet \citep{SY22, RogersSchlichting2024} also highlights the potential for an evolution in atmospheric composition as hydrogen-rich material is removed via escape. This work establishes that such complexities may impact not just the interpretations of spectroscopic data from sub-Neptunes, but the evolution of the entire small planet population. Furthermore, incorporating metallicity evolution into existing thermal and mass loss evolution models could place constraints on plausible compositional histories of sub-Neptunes, similar to how such models have previously constrained the core mass and entropy distributions on a population level.

\subsection{Inclusion of photo-evaporation}\label{sec:photoevap}
To isolate the physics, in this work we do not include XUV radiation from the host star, and we therefore do not capture the effects of photo-evaporation. Photo-evaporation has long been thought to play an important role in the evolution of super-Earths and sub-Neptunes \citep[e.g.][]{Murray-Clay2009, OwenJackson12, LopezFortney2013, OwenWu13}. In all likelihood, core-powered mass loss and photo-evaporation both play a role in the evolution of small planets. In fact, the outflow may transition from being bolometrically-driven to XUV-driven as the planet contracts and loses mass \citep{OwenSchlichting2024}. However, the specifics of such a transition need to be determined using hydrodynamic simulations. By modeling core-powered mass loss self-consistently in a hydrodynamic, radiative-transfer model, this work lays the essential groundwork toward a combined hydrodynamic model of core-powered and photo-evaporative loss. Since \textsc{aiolos} can self-consistently incorporate XUV-driven outflows \citep{SchulikBooth2023}, in future work we plan to include XUV heating in the models presented here, thus probing the interplay between the two mass-loss mechanisms proposed to form the radius valley.

\section{Conclusions}\label{sec:conclusions}
In this work, we model core-powered mass loss using a self-consistent radiative hydrodynamic code for the first time. We find that different opacity assumptions, specifically varying the ratio between the opacity to incident stellar and the opacity to outgoing thermal radiation, $\gamma$, changes the thermal and density profile in the upper atmosphere. Since the mass loss rate is sensitive to the temperature and density at the sonic radius, these changes in density structure affect the mass loss rate. We find that over a range of physically plausible values of $\gamma$, core-powered mass loss rates can vary by orders of magnitude. For a $5 M_\oplus$ planet at $T_\mathrm{eq} = 1000$~K and an RCB radius $\sim 1.2 R_\mathrm{c}$, the mass loss rate can vary from $10^{-5} \times$ the isothermal rate for $\gamma = 0.03$ to $10^5 \times$ the isothermal rate for $\gamma=30$. The differences between $\gamma$ values decrease for more inflated planet radii, converging to an analytic value given by approximating the wind as an isothermal Parker-type outflow at temperature $T = T_\mathrm{in} = T_\mathrm{eq}/2^{1/4}$. We incorporate these mass loss rates into a mass and thermal evolution model and show that varying $\gamma$ leads to variations in the evolution and expected final state of the atmosphere. Depending on the planet's parameters, including planet mass, atmospheric mass, and equilibrium temperature, the choice of $\gamma$ can make the difference between whether a planet retains a significant hydrogen atmosphere after core-powered mass loss. Preservation of more massive primordial atmospheres is favored for lower $\gamma$ values due to the lower temperatures in the upper atmosphere.

These results are relevant for predicting mass loss rates for known exoplanets. For $\gamma \approx 1$, an isothermal outflow is a good approximation of our hydrodynamic results, but at a temperature slightly lower than the equilibrium temperature. But if $\gamma$ is independently known to be very different than 1, either via retrievals of the temperature profile or based on the atmospheric compositions inferred from measurements and/or planet-wide chemical evolution models \citep[e.g.][]{MS22}, then we predict that the mass loss rate will be substantially different than the isothermal approximation. These considerations are most important for planets which have already contracted, allowing more of the outer radiative region at $T \sim T_\mathrm{out}$ to enter the sub-sonic region. For inflated planets, the isothermal formula well-approximates our hydrodynamic results to within a factor of 10 regardless of $\gamma$.

Inverting this problem, these results indicate an avenue to directly constrain the opacity ratio $\gamma$ via detections of atmospheric escape. Knowledge of the relative opacities in the upper atmosphere could then allow for inferences the composition of the upper atmosphere. Moreover, by modeling core-powered mass loss in a hydrodynamic radiative-transfer code, this work paves the way for the self-consistent combination of core-powered and photo-evaporative escape.

\section*{Acknowledgements}
This research has been supported by NASA's Exoplanet Research Program (XRP) under grant number 80NSSC21K0392. WM has also been supported by a UCLA Dissertation Year Fellowship. JEO is supported by a Royal Society University Research Fellowship. JEO and MS have received funding from the European Research Council (ERC) under the European Union’s Horizon 2020 research and innovation programme (grant agreement no. 853022, PEVAP). We thank the anonymous reviewer for their comments, which improved the clarity of the manuscript.

\vspace{5mm}
\software{\textsc{numpy} \citep{numpy},
          \textsc{matplotlib} \citep{Matplotlib},
          \textsc{scipy} \citep{scipy}}






\appendix

\section{Analytic toy model to reproduce trends in $\dot{M}$}\label{sec:app_toy}
In order to gain a simple quantitative understanding of the results in Fig.~\ref{fig:Mdot_vs_RCB}, we present a simple two-temperature toy model of escape rates.
The mass-loss rates are computed by approximating the density at the sonic point with the hydrostatic density, constructed on top of a two-temperature piecewise profile. The temperature in the inner region is $T_\mathrm{in} - 100 \log_{10}(\gamma)$, to empirically capture the inverse scaling of the inner region temperature with $\gamma$ we describe in Sec.~\ref{sec:gamma_results}. The temperature in the outer region is $T_\mathrm{out} \equiv T_\mathrm{eq} (\gamma/4)^{1/4}$, as defined in Sec.~\ref{sec:method_structure_theory}.
The transition between the two temperatures is placed at the radius corresponding to $\tau_\mathrm{P, \odot}=1$. This optical surface is calculated self-consistently from the optical depth of the lower atmosphere.

\begin{figure*}
   \centering
   \includegraphics[width=1.0\textwidth]{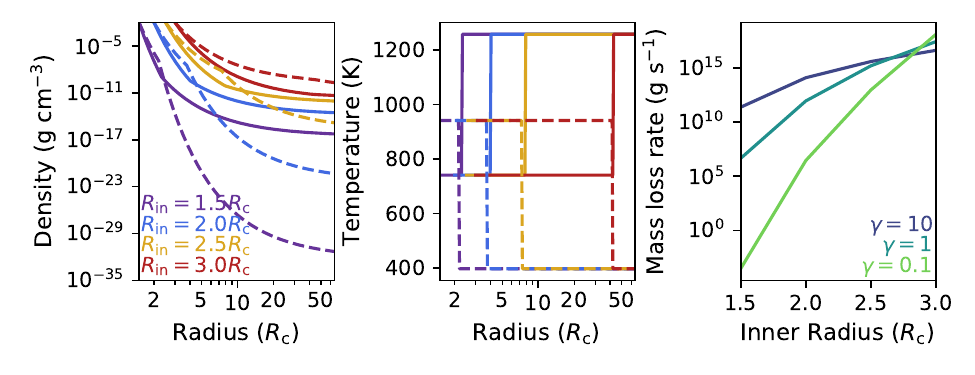}
    \caption{Demonstration of the toy model of a piecewise temperature structure with hydrostatic density profiles. In the left and center panels are radial profiles of density (left) and temperature (center). Each color represents an inner radius at which the density is equal to a fixed value. Dashed lines are $\gamma = 0.1$, while solid lines are $\gamma =10$. On the right panel are the mass loss rates for each $\gamma$ value (colors) as functions of the inner boundary radius. These mass loss rates reproduce the trends seen in Fig.~\ref{fig:Mdot_vs_RCB}. Input parameters into the toy model are $M_\mathrm{p}=5 M_\oplus$ and $T_\mathrm{eq}=1000$~K.}
    \label{fig:appendix_toymodel}
\end{figure*}

In Fig.~\ref{fig:appendix_toymodel} we show the results of our toy model. In the left and center panels we show the density and temperature profiles, respectively, as functions of radii. Each color represents a different starting inner radius for the model, at which point the density is fixed at $\rho = 10^{-2}$ g/ cm$^{-3}$. The dashed lines represent a $\gamma = 0.1$ model, while the solid lines represent a $\gamma = 10$ model. On the right panel, we show the resultant mass loss rates for three different values of $\gamma$ as functions of the inner boundary radius.

By comparison with Figs.~\ref{fig:profile_comp} and Fig.~\ref{fig:Mdot_vs_RCB}, we see that our toy model qualitatively reproduces the trends in $\dot{M}(R)$ and $\dot{M}(\gamma)$. The two temperature structures enforce different scale heights on the density profiles, causing the low $\gamma$ cases to fall off more slowly with radius in the inner region and more quickly in the outer region. For the deeper inner radii, the outer temperature dominates, leading to much lower mass loss rates for low values of $\gamma$. But as the radius increases, so too does the transition point between $T_\mathrm{in}$ and $T_\mathrm{out}$, leading to the region governed by $T_\mathrm{in}$ playing more of a role. At sufficiently large radii, the mass loss rate is set mostly by the inner temperature, and thus the mass loss rates for low values of $\gamma$ become larger than the high $\gamma$ cases, reproducing the behavior of our numerical models, e.g. Fig.~\ref{fig:Mdot_vs_RCB}.

\section{Tabulated mass loss rates}\label{sec:app_tabulated}
To ensure this work's utility for the greater scientific community, we include tabulated mass loss rates for the cases we present in Section~\ref{sec:results} as functions of $\gamma$ and the transit radius. The tables are available as machine-readable csv files in the supplementary data. To demonstrate their content, we include the first entries in the $M=5 M_\oplus$, $T_\mathrm{eq} = 1000$~K table in Table~\ref{tab:example_table}.

\begin{table*}
    \centering
    \begin{tabular}{ccccc}
        $\gamma$ & $\kappa_\mathrm{P, therm}$ (cm$^2$\ g$^{-1}$) & $\kappa_\mathrm{P, \odot}$ (cm$^2$\ g$^{-1}$) & $R_\mathrm{trans}$ (cm) & $\dot{M}$ (g\ s$^{-1}$) \\
        \hline
        $3 \times 10^{-2}$ & 7.5 & $2.25 \times 10^{-1}$ & $1.266 \times 10^9$ & $3.735 \times 10^{-8}$ \\
        $3 \times 10^{-2}$ & 7.5 & $2.25 \times 10^{-1}$ & $1.400 \times 10^9$ & $4.362 \times 10^{-5}$ \\
         & &...& & \\
    \end{tabular}
    \caption{Tabulated Mass Loss Rates}
    \tablecomments{The first entries in the tabulated mass loss rates tables, for a planet with $M_\mathrm{p}=5 M_\oplus$ and $T_\mathrm{eq} = 1000$~K. Each row corresponds to a unique \textsc{aiolos} run. The first column is the ratio of the thermal and stellar opacities, $\gamma$, used in the run, while the second and third columns are the specific Planck opacity to thermal, $\kappa_\mathrm{P, therm}$, and stellar, $\kappa_\mathrm{P, \odot}$ radiation, both in cm$^2$~g$^{-1}$. The fourth column is the transit radius, $R_\mathrm{trans}$, calculated according to Eq.~\ref{eq:transit_integral}, in cm. The fifth column is the resultant mass loss rate, $\dot{M}$, in g~s$^{-1}$. These tables can be interpolated to calculate the bolometrically-driven mass loss rate as functions of opacity ratio, observed radius, core mass, and equilibrium temperature. A machine-readable version of this table and those for other combinations of planet mass and equilibrium temperature are available as Supplementary Data.}
    \label{tab:example_table}
\end{table*}

Each file corresponds to a particular planet mass, $M_\mathrm{p}$, and equilibrium temperature, $T_\mathrm{eq}$, calculated via Eq.~\ref{eq:Teq} and is named \texttt{aiolos\_R\_Mdot\_MX\_TY.csv}, where \texttt{X} is the planet mass in Earth masses and \texttt{Y} is the equilibrium temperature in K. Each row corresponds to a unique \textsc{aiolos} run, with a particular choice of $\gamma$ and radius, and the resultant mass loss rate we obtain. The first column is the ratio of the thermal and stellar opacities, $\gamma$, used in the run. The second and third columns are the specific Planck opacities to thermal, $\kappa_\mathrm{P, therm}$, and stellar, $\kappa_\mathrm{P, \odot}$ radiation, both in cm$^2$\ g$^{-1}$, such that the ratio of the third to the second column yields the first column. The fourth column is the transit radius, $R_\mathrm{trans}$, calculated according to Eq.~\ref{eq:transit_integral}, in cm. The fifth column is the resultant mass loss rate from our \textsc{aiolos} simulations, $\dot{M}$, in g\ s$^{-1}$. We include tables for three different planet masses, $M_\mathrm{p} = 3$, $5$, and $10 M_\oplus$, at $T_\mathrm{eq} = 1000$~K, and additional tables for $M_\mathrm{p} = 5 M_\oplus$ at $T_\mathrm{eq} = 750$ and $1250$~K. Each table contains opacity ratios $\gamma = 0.03, 0.1, 0.3, 1, 3, 10$, and $30$. These tables can be interpolated to calculate the bolometrically-driven mass loss rate as functions of opacity ratio, transit radius, core mass, and equilibrium temperature for observed planets.

\bibliography{cpopa} 
\bibliographystyle{aasjournal}

\end{document}